\documentclass[
aps,prl,reprint,superscriptaddress,longbibliography,floatfix,nobibnotes
]{revtex4-2}

\usepackage{amsmath,amssymb,bm,mathtools}
\usepackage{graphicx}
\usepackage{xcolor}
\usepackage{tikz}
\usetikzlibrary{calc,positioning}
\usepackage[colorlinks=true,linkcolor=blue,citecolor=blue,urlcolor=blue]{hyperref}

\allowdisplaybreaks

\newcommand{\kk}{\bm{k}}
\newcommand{\qq}{\bm{q}}
\newcommand{\EE}{\bm{E}}
\newcommand{\del}{\partial}
\newcommand{\cL}{\mathcal{L}}
\newcommand{\cW}{\mathcal{W}}
\newcommand{\cB}{\mathcal{B}}
\newcommand{\cS}{\mathcal{S}}
\newcommand{\tr}{\operatorname{Tr}}
\newcommand{\BZ}{\mathrm{BZ}}
\newcommand{\conn}{\mathrm{conn}}
\newcommand{\disc}{\mathrm{disc}}
\newcommand{\Ree}{\operatorname{Re}}
\newcommand{\Imm}{\operatorname{Im}}
\newcommand{\Dtilde}{\widetilde{\Delta}}

\tikzset{
  skelarr/.style={gray!70,densely dotted,line width=0.95pt},
  bare/.style={line width=0.62pt},
  bw/.style={line width=1.70pt},
  poleii/.style={double,double distance=1.25pt,line width=0.45pt,line cap=butt},
  poleiii/.style={double,double distance=2.05pt,line width=0.38pt,line cap=butt,
    postaction={draw,line width=0.38pt,line cap=butt}},
  bwpoleii/.style={double,double distance=1.75pt,line width=0.80pt,line cap=butt},
  vertex/.style={circle,fill=black,inner sep=1.45pt},
  covmark/.style={circle,draw,fill=white,line width=0.85pt,inner sep=1.75pt},
  field/.style={dashed,line width=0.88pt},
  lab/.style={font=\scriptsize},
  tiny/.style={font=\tiny},
  qedge/.style={draw=gray!70,densely dotted,line width=0.95pt},
  qvsolid/.style={circle,fill=black,inner sep=1.6pt},
  qvopen/.style={circle,draw,fill=white,line width=0.9pt,inner sep=1.9pt},
  qlab/.style={font=\scriptsize}
}

\newcommand{\SkelQfourConnZero}{%
\begin{tikzpicture}[x=0.95cm,y=0.95cm,baseline=-0.1cm]
  \coordinate (v1) at (90:1.10);
  \coordinate (v2) at (18:1.10);
  \coordinate (v3) at (306:1.10);
  \coordinate (v4) at (234:1.10);
  \coordinate (v5) at (162:1.10);

  \draw[qedge] (v1)--(v2)--(v3)--(v4)--(v5)--cycle;

  \node[qvsolid] (n1) at (v1) {};
  \node[qvsolid] (n2) at (v2) {};
  \node[qvsolid] (n3) at (v3) {};
  \node[qvsolid] (n4) at (v4) {};
  \node[qvsolid] (n5) at (v5) {};

  \node[qlab] at ($(n1)+(0,0.34)$) {$\alpha$};
  \node[qlab] at ($(n2)+(0.40,0.12)$) {$\beta_1$};
  \node[qlab] at ($(n3)+(0.28,-0.33)$) {$\beta_2$};
  \node[qlab] at ($(n4)+(-0.28,-0.33)$) {$\beta_3$};
  \node[qlab] at ($(n5)+(-0.40,0.12)$) {$\beta_4$};
\end{tikzpicture}%
}

\newcommand{\SkelQfourConnOne}{%
\begin{tikzpicture}[x=0.95cm,y=0.95cm,baseline=-0.1cm]
  \coordinate (v1) at (-0.95,0.95);
  \coordinate (v2) at ( 0.95,0.95);
  \coordinate (v3) at ( 0.95,-0.95);
  \coordinate (v4) at (-0.95,-0.95);

  \draw[qedge] (v1)--(v2)--(v3)--(v4)--cycle;

  \node[qvsolid] (n1) at (v1) {};
  \node[qvsolid] (n2) at (v2) {};
  \node[qvsolid] (n3) at (v3) {};
  \node[qvopen]  (n4) at (v4) {};

  \node[qlab] at ($(n1)+(-0.28,0.34)$) {$\alpha$};
  \node[qlab] at ($(n2)+(0.28,0.34)$) {$\beta_2$};
  \node[qlab] at ($(n3)+(0.32,-0.34)$) {$\beta_3$};
  \node[qlab] at ($(n4)+(-0.04,-0.46)$) {$\scriptstyle (\beta_1|\beta_4)$};
\end{tikzpicture}%
}

\newcommand{\SkelQfourConnTwoA}{%
\begin{tikzpicture}[x=0.95cm,y=0.95cm,baseline=-0.1cm]
  \coordinate (v1) at (0,1.00);
  \coordinate (v2) at (1.05,-0.72);
  \coordinate (v3) at (-1.05,-0.72);

  \draw[qedge] (v1)--(v2)--(v3)--cycle;

  \node[qvsolid] (n1) at (v1) {};
  \node[qvsolid] (n2) at (v2) {};
  \node[qvopen]  (n3) at (v3) {};

  \node[qlab] at ($(n1)+(0,0.34)$) {$\alpha$};
  \node[qlab] at ($(n2)+(0.38,-0.30)$) {$\beta_3$};
  \node[qlab] at ($(n3)+(-0.02,-0.46)$) {$\scriptstyle (\beta_1\beta_2|\beta_4)$};
\end{tikzpicture}%
}

\newcommand{\SkelQfourConnTwoB}{%
\begin{tikzpicture}[x=0.95cm,y=0.95cm,baseline=-0.1cm]
  \coordinate (v1) at (0,1.00);
  \coordinate (v2) at (1.05,-0.72);
  \coordinate (v3) at (-1.05,-0.72);

  \draw[qedge] (v1)--(v2)--(v3)--cycle;

  \node[qvsolid] (n1) at (v1) {};
  \node[qvopen]  (n2) at (v2) {};
  \node[qvopen]  (n3) at (v3) {};

  \node[qlab] at ($(n1)+(0,0.34)$) {$\alpha$};
  \node[qlab] at ($(n2)+(0.32,-0.34)$) {$\scriptstyle (\beta_1|\beta_3)$};
  \node[qlab] at ($(n3)+(-0.32,-0.34)$) {$\scriptstyle (\beta_2|\beta_4)$};
\end{tikzpicture}%
}

\newcommand{\SkelQfourConnThree}{%
\begin{tikzpicture}[x=0.95cm,y=0.95cm,baseline=-0.1cm]
  \coordinate (L) at (-1.12,0);
  \coordinate (R) at ( 1.12,0);

  \draw[qedge] (L) .. controls (-0.60,0.55) and (0.60,0.55) .. (R);
  \draw[qedge] (L) .. controls (-0.60,-0.55) and (0.60,-0.55) .. (R);

  \node[qvsolid] (n1) at (L) {};
  \node[qvopen]  (n2) at (R) {};

  \node[qlab] at ($(n1)+(-0.34,0.26)$) {$\alpha$};
  \node[qlab] at ($(n2)+(1.00,0.34)$) {$\scriptstyle (\beta_1\beta_2\beta_3|\beta_4)$};
\end{tikzpicture}%
}

\newcommand{\FieldUp}[3]{\draw[field] (#1,#2)--(#1,#3);}
\newcommand{\SkelBubble}[1]{%
\begin{tikzpicture}[x=1cm,y=1cm,baseline=-0.35ex]
  \coordinate (L) at (-0.82,0); \coordinate (R) at (0.82,0);
  \draw[skelarr] (L) to[bend left=43] (R);
  \draw[skelarr] (R) to[bend left=43] (L);
  \node[vertex] at (L) {}; \node[vertex] at (R) {};
  \node[lab,above] at (-0.82,0.12) {$\alpha$};
  \node[lab,above] at (0.82,0.12) {$\beta_1$};
  #1
\end{tikzpicture}}

\newcommand{\SkelTriangle}[1]{%
\begin{tikzpicture}[x=1cm,y=1cm,baseline=-0.45ex]
  \coordinate (A) at (90:0.78); \coordinate (B) at (210:0.78); \coordinate (C) at (330:0.78);
  \draw[skelarr] (A)--(B); \draw[skelarr] (B)--(C); \draw[skelarr] (C)--(A);
  \foreach \P in {A,B,C}{\node[vertex] at (\P) {};}
  \node[lab,above] at (A) {$\alpha$};
  \node[lab,left] at (B) {$\beta_1$};
  \node[lab,right] at (C) {$\beta_2$};
  #1
\end{tikzpicture}}

\newcommand{\SkelSquare}[1]{%
\begin{tikzpicture}[x=1cm,y=1cm,baseline=-0.45ex]
  \coordinate (A) at (-0.70,0.70); \coordinate (B) at (0.70,0.70); \coordinate (C) at (0.70,-0.70); \coordinate (D) at (-0.70,-0.70);
  \draw[skelarr] (A)--(B); \draw[skelarr] (B)--(C); \draw[skelarr] (C)--(D); \draw[skelarr] (D)--(A);
  \foreach \P in {A,B,C,D}{\node[vertex] at (\P) {};}
  \node[lab,above left] at (A) {$\alpha$};
  \node[lab,above right] at (B) {$\beta_1$};
  \node[lab,below right] at (C) {$\beta_2$};
  \node[lab,below left] at (D) {$\beta_3$};
  #1
\end{tikzpicture}}

\newcommand{\SkelCovBubbleOne}{%
\begin{tikzpicture}[x=1cm,y=1cm,baseline=-0.35ex]
  \coordinate (L) at (-0.82,0); \coordinate (R) at (0.82,0);
  \draw[skelarr] (L) to[bend left=43] (R);
  \draw[skelarr] (R) to[bend left=43] (L);
  \node[vertex] at (L) {}; \node[covmark] at (R) {};
  \node[lab,above] at (-0.82,0.12) {$\alpha$};
  \node[lab,above] at (0.98,0.20) {$\scriptstyle(\beta_1|\beta_2)$};
\end{tikzpicture}}

\newcommand{\SkelCovTriangleOne}{%
\begin{tikzpicture}[x=1cm,y=1cm,baseline=-0.45ex]
  \coordinate (A) at (90:0.78); \coordinate (B) at (210:0.78); \coordinate (C) at (330:0.78);
  \draw[skelarr] (A)--(B); \draw[skelarr] (B)--(C); \draw[skelarr] (C)--(A);
  \node[vertex] at (A) {}; \node[vertex] at (B) {}; \node[covmark] at (C) {};
  \node[lab,above] at (A) {$\alpha$};
  \node[lab,left] at (B) {$\beta_1$};
  \node[lab,right] at (C) {$\scriptstyle(\beta_2|\beta_3)$};
\end{tikzpicture}}

\newcommand{\SkelCovBubbleTwo}{%
\begin{tikzpicture}[x=1cm,y=1cm,baseline=-0.35ex]
  \coordinate (L) at (-0.82,0); \coordinate (R) at (0.82,0);
  \draw[skelarr] (L) to[bend left=43] (R);
  \draw[skelarr] (R) to[bend left=43] (L);
  \node[vertex] at (L) {}; \node[covmark] at (R) {};
  \node[lab,above] at (-0.82,0.12) {$\alpha$};
  \node[lab,above] at (1.02,0.22) {$\scriptstyle(\beta_1\beta_2|\beta_3)$};
\end{tikzpicture}}

\newcommand{\EnergyBubbleBW}{%
\begin{tikzpicture}[x=1cm,y=1cm,baseline=-0.40ex]
  \coordinate (L) at (-0.90,0); \coordinate (R) at (0.90,0);
  \draw[bw] (L) to[bend left=43] (R);
  \draw[bw] (R) to[bend left=43] (L);
  \node[vertex] at (L) {}; \node[vertex] at (R) {};
  \FieldUp{-0.90}{0.08}{0.95}; \FieldUp{0.90}{0.08}{0.95};
  \node[lab] at (-0.90,1.12) {$a$}; \node[lab] at (0.90,1.12) {$b$};
\end{tikzpicture}}

\newcommand{\EnergyBubbleBare}{%
\begin{tikzpicture}[x=1cm,y=1cm,baseline=-0.40ex]
  \coordinate (L) at (-0.90,0); \coordinate (R) at (0.90,0);
  \draw[bare] (L) to[bend left=43] (R);
  \draw[bare] (R) to[bend left=43] (L);
  \node[vertex] at (L) {}; \node[vertex] at (R) {};
  \FieldUp{-0.90}{0.08}{0.95}; \FieldUp{0.90}{0.08}{0.95};
  \node[lab] at (-0.90,1.12) {$a$}; \node[lab] at (0.90,1.12) {$b$};
\end{tikzpicture}}

\newcommand{\EnergyBubbleDoubleBare}{%
\begin{tikzpicture}[x=1cm,y=1cm,baseline=-0.40ex]
  \coordinate (L) at (-0.90,0); \coordinate (R) at (0.90,0);
  \draw[poleii] (L) to[bend left=43] (R);
  \draw[bare] (R) to[bend left=43] (L);
  \node[vertex] at (L) {}; \node[vertex] at (R) {};
  \FieldUp{-0.90}{0.08}{0.95}; \FieldUp{0.90}{0.08}{0.95};
  \node[lab] at (-0.90,1.12) {$a$}; \node[lab] at (0.90,1.12) {$b$};
\end{tikzpicture}}

\newcommand{\EnergyBubbleTripleBare}{%
\begin{tikzpicture}[x=1cm,y=1cm,baseline=-0.40ex]
  \coordinate (L) at (-0.90,0); \coordinate (R) at (0.90,0);
  \draw[poleiii] (L) to[bend left=43] (R);
  \draw[bare] (R) to[bend left=43] (L);
  \node[vertex] at (L) {}; \node[vertex] at (R) {};
  \FieldUp{-0.90}{0.08}{0.95}; \FieldUp{0.90}{0.08}{0.95};
  \node[lab] at (-0.90,1.12) {$a$}; \node[lab] at (0.90,1.12) {$b$};
\end{tikzpicture}}

\newcommand{\EnergyTriangleBW}{%
\begin{tikzpicture}[x=1cm,y=1cm,baseline=-0.46ex]
  \coordinate (A) at (90:0.83); \coordinate (B) at (210:0.83); \coordinate (C) at (330:0.83);
  \draw[bw] (A)--(B); \draw[bw] (B)--(C); \draw[bw] (C)--(A);
  \draw[field] (A)--(0,1.65); \draw[field] (B)--(-0.72,-1.45); \draw[field] (C)--(0.72,-1.45);
  \foreach \P in {A,B,C}{\node[vertex] at (\P) {};}
  \node[lab] at (0,1.82) {$a$}; \node[lab] at (-0.72,-1.62) {$b$}; \node[lab] at (0.72,-1.62) {$c$};
\end{tikzpicture}}

\newcommand{\EnergyTriangleBare}{%
\begin{tikzpicture}[x=1cm,y=1cm,baseline=-0.46ex]
  \coordinate (A) at (90:0.83); \coordinate (B) at (210:0.83); \coordinate (C) at (330:0.83);
  \draw[bare] (A)--(B); \draw[bare] (B)--(C); \draw[bare] (C)--(A);
  \draw[field] (A)--(0,1.65); \draw[field] (B)--(-0.72,-1.45); \draw[field] (C)--(0.72,-1.45);
  \foreach \P in {A,B,C}{\node[vertex] at (\P) {};}
  \node[lab] at (0,1.82) {$a$}; \node[lab] at (-0.72,-1.62) {$b$}; \node[lab] at (0.72,-1.62) {$c$};
\end{tikzpicture}}

\newcommand{\EnergyTriangleDoubleBare}{%
\begin{tikzpicture}[x=1cm,y=1cm,baseline=-0.46ex]
  \coordinate (A) at (90:0.83); \coordinate (B) at (210:0.83); \coordinate (C) at (330:0.83);
  \draw[poleii] (A)--(B); \draw[bare] (B)--(C); \draw[bare] (C)--(A);
  \draw[field] (A)--(0,1.65); \draw[field] (B)--(-0.72,-1.45); \draw[field] (C)--(0.72,-1.45);
  \foreach \P in {A,B,C}{\node[vertex] at (\P) {};}
  \node[lab] at (0,1.82) {$a$}; \node[lab] at (-0.72,-1.62) {$b$}; \node[lab] at (0.72,-1.62) {$c$};
\end{tikzpicture}}

\newcommand{\EnergySquareBW}{%
\begin{tikzpicture}[x=1cm,y=1cm,baseline=-0.48ex]
  \coordinate (A) at (-0.78,0.78); \coordinate (B) at (0.78,0.78); \coordinate (C) at (0.78,-0.78); \coordinate (D) at (-0.78,-0.78);
  \draw[bw] (A)--(B); \draw[bw] (B)--(C); \draw[bw] (C)--(D); \draw[bw] (D)--(A);
  \foreach \P in {A,B,C,D}{\node[vertex] at (\P) {};}
  \FieldUp{-0.78}{0.86}{1.48}; \FieldUp{0.78}{0.86}{1.48}; \FieldUp{0.78}{-0.86}{-1.48}; \FieldUp{-0.78}{-0.86}{-1.48};
  \node[lab] at (-0.78,1.65) {$a$}; \node[lab] at (0.78,1.65) {$b$}; \node[lab] at (0.78,-1.65) {$c$}; \node[lab] at (-0.78,-1.65) {$d$};
\end{tikzpicture}}

\newcommand{\EnergySquareBare}{%
\begin{tikzpicture}[x=1cm,y=1cm,baseline=-0.48ex]
  \coordinate (A) at (-0.78,0.78); \coordinate (B) at (0.78,0.78); \coordinate (C) at (0.78,-0.78); \coordinate (D) at (-0.78,-0.78);
  \draw[bare] (A)--(B); \draw[bare] (B)--(C); \draw[bare] (C)--(D); \draw[bare] (D)--(A);
  \foreach \P in {A,B,C,D}{\node[vertex] at (\P) {};}
  \FieldUp{-0.78}{0.86}{1.48}; \FieldUp{0.78}{0.86}{1.48}; \FieldUp{0.78}{-0.86}{-1.48}; \FieldUp{-0.78}{-0.86}{-1.48};
  \node[lab] at (-0.78,1.65) {$a$}; \node[lab] at (0.78,1.65) {$b$}; \node[lab] at (0.78,-1.65) {$c$}; \node[lab] at (-0.78,-1.65) {$d$};
\end{tikzpicture}}

\newcommand{\EnergySquareDoubleBare}{%
\begin{tikzpicture}[x=1cm,y=1cm,baseline=-0.48ex]
  \coordinate (A) at (-0.78,0.78); \coordinate (B) at (0.78,0.78); \coordinate (C) at (0.78,-0.78); \coordinate (D) at (-0.78,-0.78);
  \draw[poleii] (A)--(B); \draw[bare] (B)--(C); \draw[bare] (C)--(D); \draw[bare] (D)--(A);
  \foreach \P in {A,B,C,D}{\node[vertex] at (\P) {};}
  \FieldUp{-0.78}{0.86}{1.48}; \FieldUp{0.78}{0.86}{1.48}; \FieldUp{0.78}{-0.86}{-1.48}; \FieldUp{-0.78}{-0.86}{-1.48};
  \node[lab] at (-0.78,1.65) {$a$}; \node[lab] at (0.78,1.65) {$b$}; \node[lab] at (0.78,-1.65) {$c$}; \node[lab] at (-0.78,-1.65) {$d$};
\end{tikzpicture}}

\newcommand{\NormBubbleBW}{%
\begin{tikzpicture}[x=1cm,y=1cm,baseline=-0.40ex]
  \coordinate (L) at (-0.90,0); \coordinate (R) at (0.90,0);
  \draw[bwpoleii] (L) to[bend left=43] (R);
  \draw[bw] (R) to[bend left=43] (L);
  \node[vertex] at (L) {}; \node[vertex] at (R) {};
  \FieldUp{-0.90}{0.08}{0.95}; \FieldUp{0.90}{0.08}{0.95};
\end{tikzpicture}}

\newcommand{\NormTriangleBW}{%
\begin{tikzpicture}[x=1cm,y=1cm,baseline=-0.46ex]
  \coordinate (A) at (90:0.83); \coordinate (B) at (210:0.83); \coordinate (C) at (330:0.83);
  \draw[bwpoleii] (A)--(B); \draw[bw] (B)--(C); \draw[bw] (C)--(A);
  \draw[field] (A)--(0,1.55); \draw[field] (B)--(-0.72,-1.35); \draw[field] (C)--(0.72,-1.35);
  \foreach \P in {A,B,C}{\node[vertex] at (\P) {};}
\end{tikzpicture}}

\newcommand{\NormSquareBW}{%
\begin{tikzpicture}[x=1cm,y=1cm,baseline=-0.48ex]
  \coordinate (A) at (-0.78,0.78); \coordinate (B) at (0.78,0.78); \coordinate (C) at (0.78,-0.78); \coordinate (D) at (-0.78,-0.78);
  \draw[bwpoleii] (A)--(B); \draw[bw] (B)--(C); \draw[bw] (C)--(D); \draw[bw] (D)--(A);
  \foreach \P in {A,B,C,D}{\node[vertex] at (\P) {};}
  \FieldUp{-0.78}{0.86}{1.42}; \FieldUp{0.78}{0.86}{1.42}; \FieldUp{0.78}{-0.86}{-1.42}; \FieldUp{-0.78}{-0.86}{-1.42};
\end{tikzpicture}}

\newcommand{\ConnBubbleBW}{%
\begin{tikzpicture}[x=1cm,y=1cm,baseline=-0.40ex]
  \coordinate (L) at (-0.90,0); \coordinate (R) at (0.90,0);
  \draw[bw] (L) to[bend left=43] (R); \draw[bw] (R) to[bend left=43] (L);
  \node[vertex] at (L) {}; \node[vertex] at (R) {};
  \FieldUp{0.90}{0.08}{0.95};
  \node[lab] at (-1.08,0.28) {$c$}; \node[lab] at (0.90,1.12) {$a$};
\end{tikzpicture}}

\newcommand{\ConnTriangleBW}{%
\begin{tikzpicture}[x=1cm,y=1cm,baseline=-0.46ex]
  \coordinate (A) at (90:0.83); \coordinate (B) at (210:0.83); \coordinate (C) at (330:0.83);
  \draw[bw] (A)--(B); \draw[bw] (B)--(C); \draw[bw] (C)--(A);
  \draw[field] (B)--(-0.72,-1.45); \draw[field] (C)--(0.72,-1.45);
  \node[vertex] at (A) {}; \node[vertex] at (B) {}; \node[vertex] at (C) {};
  \node[lab] at (0,1.18) {$c$}; \node[lab] at (-0.72,-1.62) {$a$}; \node[lab] at (0.72,-1.62) {$b$};
\end{tikzpicture}}

\newcommand{\ConnCovBubbleBW}{%
\begin{tikzpicture}[x=1cm,y=1cm,baseline=-0.40ex]
  \coordinate (L) at (-0.95,0); \coordinate (R) at (0.95,0);
  \draw[bw] (L) to[bend left=43] (R); \draw[bw] (R) to[bend left=43] (L);
  \node[vertex] at (L) {}; \node[covmark] at (R) {};
  \FieldUp{-0.95}{0.08}{0.95};
  \node[lab] at (-0.95,1.12) {$a$};
  \node[tiny] at (1.35,0.18) {$\scriptstyle(c|b)$};
\end{tikzpicture}}

\newcommand{\ConnSquareBW}{%
\begin{tikzpicture}[x=1cm,y=1cm,baseline=-0.48ex]
  \coordinate (A) at (-0.78,0.78); \coordinate (B) at (0.78,0.78); \coordinate (C) at (0.78,-0.78); \coordinate (D) at (-0.78,-0.78);
  \draw[bw] (A)--(B); \draw[bw] (B)--(C); \draw[bw] (C)--(D); \draw[bw] (D)--(A);
  \foreach \P in {A,B,C,D}{\node[vertex] at (\P) {};}
  \FieldUp{0.78}{0.86}{1.48}; \FieldUp{0.78}{-0.86}{-1.48}; \FieldUp{-0.78}{-0.86}{-1.48};
  \node[lab] at (-1.05,1.02) {$c$}; \node[lab] at (0.78,1.65) {$a$};
  \node[lab] at (0.78,-1.65) {$b$}; \node[lab] at (-0.78,-1.65) {$d$};
\end{tikzpicture}}

\newcommand{\ConnCovTriangleBW}{%
\begin{tikzpicture}[x=1cm,y=1cm,baseline=-0.46ex]
  \coordinate (A) at (90:0.83); \coordinate (B) at (210:0.83); \coordinate (C) at (330:0.83);
  \draw[bw] (A)--(B); \draw[bw] (B)--(C); \draw[bw] (C)--(A);
  \draw[field] (B)--(-0.72,-1.45); \draw[field] (C)--(0.72,-1.45);
  \node[covmark] at (A) {}; \node[vertex] at (B) {}; \node[vertex] at (C) {};
  \node[tiny] at (0.42,1.12) {$\scriptstyle(c|d)$};
  \node[lab] at (-0.72,-1.62) {$a$}; \node[lab] at (0.72,-1.62) {$b$};
\end{tikzpicture}}

\newcommand{\ConnCovBubbleTwoBW}{%
\begin{tikzpicture}[x=1cm,y=1cm,baseline=-0.40ex]
  \coordinate (L) at (-0.95,0); \coordinate (R) at (0.95,0);
  \draw[bw] (L) to[bend left=43] (R); \draw[bw] (R) to[bend left=43] (L);
  \node[vertex] at (L) {}; \node[covmark] at (R) {};
  \FieldUp{-0.95}{0.08}{0.95};
  \node[lab] at (-0.95,1.12) {$a$};
  \node[tiny] at (1.45,0.20) {$\scriptstyle(c\,b|d)$};
\end{tikzpicture}}

\newcommand{\ConnBubbleBare}{%
\begin{tikzpicture}[x=1cm,y=1cm,baseline=-0.40ex]
  \coordinate (L) at (-0.90,0); \coordinate (R) at (0.90,0);
  \draw[bare] (L) to[bend left=43] (R); \draw[bare] (R) to[bend left=43] (L);
  \node[vertex] at (L) {}; \node[vertex] at (R) {};
  \FieldUp{0.90}{0.08}{0.95};
  \node[lab] at (-1.08,0.28) {$c$}; \node[lab] at (0.90,1.12) {$a$};
\end{tikzpicture}}

\newcommand{\ConnTriangleBare}{%
\begin{tikzpicture}[x=1cm,y=1cm,baseline=-0.46ex]
  \coordinate (A) at (90:0.83); \coordinate (B) at (210:0.83); \coordinate (C) at (330:0.83);
  \draw[bare] (A)--(B); \draw[bare] (B)--(C); \draw[bare] (C)--(A);
  \draw[field] (B)--(-0.72,-1.45); \draw[field] (C)--(0.72,-1.45);
  \node[vertex] at (A) {}; \node[vertex] at (B) {}; \node[vertex] at (C) {};
  \node[lab] at (0,1.18) {$c$}; \node[lab] at (-0.72,-1.62) {$a$}; \node[lab] at (0.72,-1.62) {$b$};
\end{tikzpicture}}

\newcommand{\ConnCovBubbleBare}{%
\begin{tikzpicture}[x=1cm,y=1cm,baseline=-0.40ex]
  \coordinate (L) at (-0.95,0); \coordinate (R) at (0.95,0);
  \draw[bare] (L) to[bend left=43] (R); \draw[bare] (R) to[bend left=43] (L);
  \node[vertex] at (L) {}; \node[covmark] at (R) {};
  \FieldUp{-0.95}{0.08}{0.95};
  \node[lab] at (-0.95,1.12) {$a$};
  \node[tiny] at (1.35,0.18) {$\scriptstyle(c|b)$};
\end{tikzpicture}}

\newcommand{\ConnCovBubbleHigherPoleBare}{%
\begin{tikzpicture}[x=1cm,y=1cm,baseline=-0.40ex]
  \coordinate (L) at (-0.95,0); \coordinate (R) at (0.95,0);
  \draw[poleii] (L) to[bend left=43] (R);
  \draw[bare] (R) to[bend left=43] (L);
  \node[vertex] at (L) {}; \node[covmark] at (R) {};
  \FieldUp{-0.95}{0.08}{0.95};
  \node[lab] at (-0.95,1.12) {$a$};
  \node[tiny] at (1.35,0.18) {$\scriptstyle(c|b)$};
\end{tikzpicture}}

\newcommand{\ConnSquareBare}{%
\begin{tikzpicture}[x=1cm,y=1cm,baseline=-0.48ex]
  \coordinate (A) at (-0.78,0.78); \coordinate (B) at (0.78,0.78); \coordinate (C) at (0.78,-0.78); \coordinate (D) at (-0.78,-0.78);
  \draw[bare] (A)--(B); \draw[bare] (B)--(C); \draw[bare] (C)--(D); \draw[bare] (D)--(A);
  \foreach \P in {A,B,C,D}{\node[vertex] at (\P) {};}
  \FieldUp{0.78}{0.86}{1.48}; \FieldUp{0.78}{-0.86}{-1.48}; \FieldUp{-0.78}{-0.86}{-1.48};
  \node[lab] at (-1.05,1.02) {$c$}; \node[lab] at (0.78,1.65) {$a$};
  \node[lab] at (0.78,-1.65) {$b$}; \node[lab] at (-0.78,-1.65) {$d$};
\end{tikzpicture}}

\newcommand{\ConnCovTriangleBare}{%
\begin{tikzpicture}[x=1cm,y=1cm,baseline=-0.46ex]
  \coordinate (A) at (90:0.83); \coordinate (B) at (210:0.83); \coordinate (C) at (330:0.83);
  \draw[bare] (A)--(B); \draw[bare] (B)--(C); \draw[bare] (C)--(A);
  \draw[field] (B)--(-0.72,-1.45); \draw[field] (C)--(0.72,-1.45);
  \node[covmark] at (A) {}; \node[vertex] at (B) {}; \node[vertex] at (C) {};
  \node[tiny] at (0.42,1.12) {$\scriptstyle(c|d)$};
  \node[lab] at (-0.72,-1.62) {$a$}; \node[lab] at (0.72,-1.62) {$b$};
\end{tikzpicture}}

\newcommand{\ConnCovBubbleTwoBare}{%
\begin{tikzpicture}[x=1cm,y=1cm,baseline=-0.40ex]
  \coordinate (L) at (-0.95,0); \coordinate (R) at (0.95,0);
  \draw[bare] (L) to[bend left=43] (R); \draw[bare] (R) to[bend left=43] (L);
  \node[vertex] at (L) {}; \node[covmark] at (R) {};
  \FieldUp{-0.95}{0.08}{0.95};
  \node[lab] at (-0.95,1.12) {$a$};
  \node[tiny] at (1.45,0.20) {$\scriptstyle(c\,b|d)$};
\end{tikzpicture}}

\newcommand{\ConnBubbleDoubleBare}{%
\begin{tikzpicture}[x=1cm,y=1cm,baseline=-0.40ex]
  \coordinate (L) at (-0.90,0); \coordinate (R) at (0.90,0);
  \draw[poleii] (L) to[bend left=43] (R);
  \draw[bare] (R) to[bend left=43] (L);
  \node[vertex] at (L) {}; \node[vertex] at (R) {};
  \FieldUp{0.90}{0.08}{0.95};
  \node[lab] at (-1.08,0.28) {$c$}; \node[lab] at (0.90,1.12) {$a$};
\end{tikzpicture}}

\newcommand{\EnergyPentagonBW}{%
\begin{tikzpicture}[x=1cm,y=1cm,baseline=-0.50ex]
  \coordinate (A) at (90:0.88); \coordinate (B) at (162:0.88); \coordinate (C) at (234:0.88); \coordinate (D) at (306:0.88); \coordinate (E) at (18:0.88);
  \draw[bw] (A)--(B); \draw[bw] (B)--(C); \draw[bw] (C)--(D); \draw[bw] (D)--(E); \draw[bw] (E)--(A);
  \foreach \P in {A,B,C,D,E}{\node[vertex] at (\P) {};}
  \draw[field] (A)--(0,1.58); \draw[field] (B)--(-1.35,0.76); \draw[field] (C)--(-1.10,-1.35); \draw[field] (D)--(1.10,-1.35); \draw[field] (E)--(1.35,0.76);
  \node[lab] at (0,1.75) {$a$}; \node[lab] at (-1.50,0.86) {$b$}; \node[lab] at (-1.25,-1.52) {$c$}; \node[lab] at (1.25,-1.52) {$d$}; \node[lab] at (1.50,0.86) {$e$};
\end{tikzpicture}}

\newcommand{\EnergyPentagonBare}{%
\begin{tikzpicture}[x=1cm,y=1cm,baseline=-0.50ex]
  \coordinate (A) at (90:0.88); \coordinate (B) at (162:0.88); \coordinate (C) at (234:0.88); \coordinate (D) at (306:0.88); \coordinate (E) at (18:0.88);
  \draw[bare] (A)--(B); \draw[bare] (B)--(C); \draw[bare] (C)--(D); \draw[bare] (D)--(E); \draw[bare] (E)--(A);
  \foreach \P in {A,B,C,D,E}{\node[vertex] at (\P) {};}
  \draw[field] (A)--(0,1.58); \draw[field] (B)--(-1.35,0.76); \draw[field] (C)--(-1.10,-1.35); \draw[field] (D)--(1.10,-1.35); \draw[field] (E)--(1.35,0.76);
  \node[lab] at (0,1.75) {$a$}; \node[lab] at (-1.50,0.86) {$b$}; \node[lab] at (-1.25,-1.52) {$c$}; \node[lab] at (1.25,-1.52) {$d$}; \node[lab] at (1.50,0.86) {$e$};
\end{tikzpicture}}

\newcommand{\EnergyPentagonDoubleBare}{%
\begin{tikzpicture}[x=1cm,y=1cm,baseline=-0.50ex]
  \coordinate (A) at (90:0.88); \coordinate (B) at (162:0.88); \coordinate (C) at (234:0.88); \coordinate (D) at (306:0.88); \coordinate (E) at (18:0.88);
  \draw[poleii] (A)--(B); \draw[bare] (B)--(C); \draw[bare] (C)--(D); \draw[bare] (D)--(E); \draw[bare] (E)--(A);
  \foreach \P in {A,B,C,D,E}{\node[vertex] at (\P) {};}
  \draw[field] (A)--(0,1.58); \draw[field] (B)--(-1.35,0.76); \draw[field] (C)--(-1.10,-1.35); \draw[field] (D)--(1.10,-1.35); \draw[field] (E)--(1.35,0.76);
\end{tikzpicture}}

\newcommand{\EnergyTriangleTripleBare}{%
\begin{tikzpicture}[x=1cm,y=1cm,baseline=-0.46ex]
  \coordinate (A) at (90:0.83); \coordinate (B) at (210:0.83); \coordinate (C) at (330:0.83);
  \draw[poleiii] (A)--(B); \draw[bare] (B)--(C); \draw[bare] (C)--(A);
  \draw[field] (A)--(0,1.65); \draw[field] (B)--(-0.72,-1.45); \draw[field] (C)--(0.72,-1.45);
  \foreach \P in {A,B,C}{\node[vertex] at (\P) {};}
\end{tikzpicture}}

\newcommand{\ConnPentagonBW}{%
\begin{tikzpicture}[x=1cm,y=1cm,baseline=-0.50ex]
  \coordinate (A) at (90:0.88); \coordinate (B) at (162:0.88); \coordinate (C) at (234:0.88); \coordinate (D) at (306:0.88); \coordinate (E) at (18:0.88);
  \draw[bw] (A)--(B); \draw[bw] (B)--(C); \draw[bw] (C)--(D); \draw[bw] (D)--(E); \draw[bw] (E)--(A);
  \foreach \P in {A,B,C,D,E}{\node[vertex] at (\P) {};}
  \draw[field] (B)--(-1.35,0.76); \draw[field] (C)--(-1.10,-1.35); \draw[field] (D)--(1.10,-1.35); \draw[field] (E)--(1.35,0.76);
  \node[lab] at (0,1.18) {$c$}; \node[lab] at (-1.50,0.86) {$a$}; \node[lab] at (-1.25,-1.52) {$b$}; \node[lab] at (1.25,-1.52) {$d$}; \node[lab] at (1.50,0.86) {$e$};
\end{tikzpicture}}

\newcommand{\ConnPentagonBare}{%
\begin{tikzpicture}[x=1cm,y=1cm,baseline=-0.50ex]
  \coordinate (A) at (90:0.88); \coordinate (B) at (162:0.88); \coordinate (C) at (234:0.88); \coordinate (D) at (306:0.88); \coordinate (E) at (18:0.88);
  \draw[bare] (A)--(B); \draw[bare] (B)--(C); \draw[bare] (C)--(D); \draw[bare] (D)--(E); \draw[bare] (E)--(A);
  \foreach \P in {A,B,C,D,E}{\node[vertex] at (\P) {};}
  \draw[field] (B)--(-1.35,0.76); \draw[field] (C)--(-1.10,-1.35); \draw[field] (D)--(1.10,-1.35); \draw[field] (E)--(1.35,0.76);
  \node[lab] at (0,1.18) {$c$}; \node[lab] at (-1.50,0.86) {$a$}; \node[lab] at (-1.25,-1.52) {$b$}; \node[lab] at (1.25,-1.52) {$d$}; \node[lab] at (1.50,0.86) {$e$};
\end{tikzpicture}}

\newcommand{\ConnCovSquareBW}{%
\begin{tikzpicture}[x=1cm,y=1cm,baseline=-0.48ex]
  \coordinate (A) at (-0.78,0.78); \coordinate (B) at (0.78,0.78); \coordinate (C) at (0.78,-0.78); \coordinate (D) at (-0.78,-0.78);
  \draw[bw] (A)--(B); \draw[bw] (B)--(C); \draw[bw] (C)--(D); \draw[bw] (D)--(A);
  \node[covmark] at (A) {}; \foreach \P in {B,C,D}{\node[vertex] at (\P) {};}
  \FieldUp{0.78}{0.86}{1.48}; \FieldUp{0.78}{-0.86}{-1.48}; \FieldUp{-0.78}{-0.86}{-1.48};
  \node[tiny] at (-1.18,1.04) {$\scriptstyle(c|e)$}; \node[lab] at (0.78,1.65) {$a$};
  \node[lab] at (0.78,-1.65) {$b$}; \node[lab] at (-0.78,-1.65) {$d$};
\end{tikzpicture}}

\newcommand{\ConnCovSquareBare}{%
\begin{tikzpicture}[x=1cm,y=1cm,baseline=-0.48ex]
  \coordinate (A) at (-0.78,0.78); \coordinate (B) at (0.78,0.78); \coordinate (C) at (0.78,-0.78); \coordinate (D) at (-0.78,-0.78);
  \draw[bare] (A)--(B); \draw[bare] (B)--(C); \draw[bare] (C)--(D); \draw[bare] (D)--(A);
  \node[covmark] at (A) {}; \foreach \P in {B,C,D}{\node[vertex] at (\P) {};}
  \FieldUp{0.78}{0.86}{1.48}; \FieldUp{0.78}{-0.86}{-1.48}; \FieldUp{-0.78}{-0.86}{-1.48};
  \node[tiny] at (-1.18,1.04) {$\scriptstyle(c|e)$};
\end{tikzpicture}}

\newcommand{\ConnCovTriTwoOneBW}{%
\begin{tikzpicture}[x=1cm,y=1cm,baseline=-0.46ex]
  \coordinate (A) at (90:0.83); \coordinate (B) at (210:0.83); \coordinate (C) at (330:0.83);
  \draw[bw] (A)--(B); \draw[bw] (B)--(C); \draw[bw] (C)--(A);
  \node[covmark] at (A) {}; \node[vertex] at (B) {}; \node[vertex] at (C) {};
  \node[tiny] at (0.48,1.12) {$\scriptstyle(c\,a|e)$};
  \draw[field] (B)--(-0.72,-1.45); \draw[field] (C)--(0.72,-1.45);
  \node[lab] at (-0.72,-1.62) {$b$}; \node[lab] at (0.72,-1.62) {$d$};
\end{tikzpicture}}

\newcommand{\ConnCovTriTwoOneBare}{%
\begin{tikzpicture}[x=1cm,y=1cm,baseline=-0.46ex]
  \coordinate (A) at (90:0.83); \coordinate (B) at (210:0.83); \coordinate (C) at (330:0.83);
  \draw[bare] (A)--(B); \draw[bare] (B)--(C); \draw[bare] (C)--(A);
  \node[covmark] at (A) {}; \node[vertex] at (B) {}; \node[vertex] at (C) {};
  \node[tiny] at (0.48,1.12) {$\scriptstyle(c\,a|e)$};
  \draw[field] (B)--(-0.72,-1.45); \draw[field] (C)--(0.72,-1.45);
\end{tikzpicture}}

\newcommand{\ConnCovTriTwoSplitBW}{%
\begin{tikzpicture}[x=1cm,y=1cm,baseline=-0.46ex]
  \coordinate (A) at (90:0.83); \coordinate (B) at (210:0.83); \coordinate (C) at (330:0.83);
  \draw[bw] (A)--(B); \draw[bw] (B)--(C); \draw[bw] (C)--(A);
  \node[covmark] at (A) {}; \node[covmark] at (B) {}; \node[vertex] at (C) {};
  \node[tiny] at (0.42,1.12) {$\scriptstyle(c|e)$};
  \node[tiny] at (-1.12,-0.70) {$\scriptstyle(a|d)$};
  \draw[field] (C)--(0.72,-1.45); \node[lab] at (0.72,-1.62) {$b$};
\end{tikzpicture}}

\newcommand{\ConnCovTriTwoSplitBare}{%
\begin{tikzpicture}[x=1cm,y=1cm,baseline=-0.46ex]
  \coordinate (A) at (90:0.83); \coordinate (B) at (210:0.83); \coordinate (C) at (330:0.83);
  \draw[bare] (A)--(B); \draw[bare] (B)--(C); \draw[bare] (C)--(A);
  \node[covmark] at (A) {}; \node[covmark] at (B) {}; \node[vertex] at (C) {};
  \node[tiny] at (0.42,1.12) {$\scriptstyle(c|e)$};
  \node[tiny] at (-1.12,-0.70) {$\scriptstyle(a|d)$};
  \draw[field] (C)--(0.72,-1.45);
\end{tikzpicture}}

\newcommand{\ConnCovBubbleThreeBW}{%
\begin{tikzpicture}[x=1cm,y=1cm,baseline=-0.40ex]
  \coordinate (L) at (-0.95,0); \coordinate (R) at (0.95,0);
  \draw[bw] (L) to[bend left=43] (R); \draw[bw] (R) to[bend left=43] (L);
  \node[vertex] at (L) {}; \node[covmark] at (R) {};
  \FieldUp{-0.95}{0.08}{0.95}; \node[lab] at (-0.95,1.12) {$a$};
  \node[tiny] at (1.55,0.22) {$\scriptstyle(c\,b\,d|e)$};
\end{tikzpicture}}

\newcommand{\ConnCovBubbleThreeBare}{%
\begin{tikzpicture}[x=1cm,y=1cm,baseline=-0.40ex]
  \coordinate (L) at (-0.95,0); \coordinate (R) at (0.95,0);
  \draw[bare] (L) to[bend left=43] (R); \draw[bare] (R) to[bend left=43] (L);
  \node[vertex] at (L) {}; \node[covmark] at (R) {};
  \FieldUp{-0.95}{0.08}{0.95};
  \node[tiny] at (1.55,0.22) {$\scriptstyle(c\,b\,d|e)$};
\end{tikzpicture}}

\newcommand{\ConnTriangleDoubleBare}{%
\begin{tikzpicture}[x=1cm,y=1cm,baseline=-0.46ex]
  \coordinate (A) at (90:0.83); \coordinate (B) at (210:0.83); \coordinate (C) at (330:0.83);
  \draw[poleii] (A)--(B); \draw[bare] (B)--(C); \draw[bare] (C)--(A);
  \draw[field] (B)--(-0.72,-1.45); \draw[field] (C)--(0.72,-1.45);
  \node[vertex] at (A) {}; \node[vertex] at (B) {}; \node[vertex] at (C) {};
  \node[lab] at (0,1.18) {$c$};
\end{tikzpicture}}

\newcommand{\NormPentagonBW}{%
\begin{tikzpicture}[x=1cm,y=1cm,baseline=-0.50ex]
  \coordinate (A) at (90:0.88); \coordinate (B) at (162:0.88); \coordinate (C) at (234:0.88); \coordinate (D) at (306:0.88); \coordinate (E) at (18:0.88);
  \draw[bwpoleii] (A)--(B); \draw[bw] (B)--(C); \draw[bw] (C)--(D); \draw[bw] (D)--(E); \draw[bw] (E)--(A);
  \foreach \P in {A,B,C,D,E}{\node[vertex] at (\P) {};}
  \draw[field] (A)--(0,1.55); \draw[field] (B)--(-1.35,0.73); \draw[field] (C)--(-1.08,-1.32); \draw[field] (D)--(1.08,-1.32); \draw[field] (E)--(1.35,0.73);
\end{tikzpicture}}

\begin{document}

\title{High Order Geometric Channels for Nonlinear Transport in Bloch Bands}

\author{Sami Farrag}
\email{farra146@umn.edu}
\affiliation{Department of Physics and Astronomy, University of Minnesota, Minneapolis, Minnesota 55455, USA}
\author{Eugene Mele}
\email{mele@upenn.edu}
\affiliation{Department of Physics and Astronomy, University of Pennsylvania, Philadelphia, Pennsylvania 19104, USA}
\author{Tony Low}
\email{tlow@umn.edu}
\affiliation{Department of Electrical \& Computer Engineering, University of Minnesota, Minneapolis, Minnesota 55455, USA}

\begin{abstract}
We develop a geometric perturbation theory of Bloch states for perturbations coupling via the interband Berry connection. The gauge-invariant Bargmann trace builds the dressed dispersion and connection order by order as a hierarchy \(Q^{(N)}\), starting with the quantum geometric tensor. Higher members encode multiband geometry beyond the quantum metric and Berry curvature. Connected amplitudes control vertex-order corrections, while strict-order corrections reduce to disconnected products. For a uniform electric field we find the fully coherent, purely geometric sector of the third-order response.
\end{abstract}

\maketitle

\section{Introduction}
\label{sec:intro}

The geometry of Bloch wavefunctions controls many electronic response coefficients. At the rank-two level, the quantum geometric tensor encodes both the quantum metric and the Berry curvature, which govern anomalous velocity, superfluid weight, Wannier localization, and topological Hall transport~\cite{ProvostVallee1980,Berry1984,Simon1983,WilczekZeePRL1984,XiaoChangNiuRMP2010,SouzaWilkensMartinPRB2000,PeottaTormaNatCommun2015,NagaosaAHE_RMP2010}. The integral of the Berry curvature over a filled band gives the Chern number and the quantized Hall conductance~\cite{TKNNPRL1982,Kohmoto1985,NiuThoulessWuPRB1985,Avron1983}, while the quantum metric controls the spread of Wannier functions and ground-state polarization fluctuations~\cite{RestaSorella1999,SouzaWilkensMartinPRB2000,MarzariVanderbiltPRB1997,MarzariRMP2012}.

Nonlinear response probes geometric data beyond the rank-two quantum geometric tensor. The central goal of this work is to formulate these higher-rank objects as a geometric perturbation theory. We consider a general perturbation with off-diagonal matrix elements
\(V_{mn}=\alpha^a A_{a,mn}\) for \(m\ne n\), where \(\alpha^a\) are real perturbation amplitudes and \(A_{a,mn}=i\langle u_m|\partial_a u_n\rangle\). The length-gauge electric-field perturbation is recovered as the special case \(\alpha^a=-eE^a\), but the construction itself does not rely on choosing an electric field. The local geometry underlying the perturbation series is generated by the three-point Bargmann invariant of band projectors~\cite{Bargmann1964,AvdoshkinPopovPRB2023}. Its Taylor expansion defines a hierarchy of tensors \(Q^{(N)}\). A covariant recursion separates the connected part of these tensors from return-through-band products, in direct analogy with a linked-cluster decomposition. The first member of the hierarchy is the quantum geometric tensor; higher members contain the rank-three and higher projective data needed for nonlinear transport.

We show at the end that, beyond the familiar Drude, Berry curvature quadrupole, and quantum metric quadrupole channels, the third-order conductivity contains two intrinsic fully coherent rank-three channels: a gradient of a three-band energy loop and a curl of the second-order connection polarizability. The former vanishes in strict two-band models, while the latter recovers the two-band result of Ref.~\cite{FangCanoPRL2024}.

\section{Local Bargmann hierarchy}
\label{sec:local}

\subsection{Bargmann three-point invariant}
\label{sec:p3}

For a nondegenerate Bloch band, the cell-periodic eigenstate
$|u_{n\kk}\rangle$ defines a map from the Brillouin zone
$M\simeq T^d$ into the projective Hilbert space of an
$N_{\rm H}$-component Bloch Hamiltonian,
\begin{equation}
\kk\in M
\;\longmapsto\;
[u_{n\kk}]\in\mathbb{CP}^{N_{\rm H}-1} .
\label{eq:bandmap}
\end{equation}
Here $[u_{n\kk}]$ denotes the ray of the Bloch state, so the map is
insensitive to the local phase choice of $|u_{n\kk}\rangle$. The local
geometry of this map is naturally organized by the three-point
Bargmann invariant
\begin{equation}
P_n^{(3)}(\kk_1,\kk_2,\kk_3)
=
\tr\!\left[
P_n(\kk_1)P_n(\kk_2)P_n(\kk_3)
\right],
\label{eq:P3}
\end{equation}
with $P_n(\kk)=|u_{n\kk}\rangle\langle u_{n\kk}|$
\cite{Bargmann1964,AvdoshkinPopovPRB2023}.
Unlike a two-point projector overlap, which only measures fidelity, the
three-point invariant also carries the Berry phase. Thus, $P_n^{(3)}$ can be
used as a generating function for the local geometry. At a base point
$\kk$ and displacement $\qq$, define the differentiated Bargmann trace
\begin{align}
\cB_{n,\alpha}(\qq;\kk)
&=\left.\del_{3\alpha}
P_n^{(3)}(\kk+\qq,\kk,\kk_3)\right|_{\kk_3=\kk}
\nonumber\\[-2pt]
&=\tr\!\left[P_n(\kk+\qq)P_n(\kk)\del_\alpha P_n(\kk)\right],
\label{eq:Bdef}
\end{align}
where $\del_{3\alpha}\equiv\partial/\partial k_{3,\alpha}$. Its Taylor
expansion around $\qq=0$ defines the local hierarchy,
\begin{equation}
\cB_{n,\alpha}(\qq;\kk)=
\sum_{N\ge1}\frac{1}{N!}
Q^{(N),n}_{\alpha;\beta_1\cdots\beta_N}(\kk)
q^{\beta_1}\cdots q^{\beta_N} .
\label{eq:Bexp}
\end{equation}
Thus
\begin{equation}
Q^{(N),n}_{\alpha;\beta_1\cdots\beta_N}
=\tr\!\left[
P_n(\del_\alpha P_n)
(\del_{\beta_1}\cdots\del_{\beta_N}P_n)
\right],
\label{eq:QNdef}
\end{equation}
with all projectors evaluated at the same momentum $\kk$. The tensor
$Q^{(N)}$ has total rank $N+1$. In particular, the rank-two first
member is the quantum geometric tensor,
$Q^{(1),n}_{\alpha;\beta}=g^n_{\alpha\beta}
-\tfrac{i}{2}\Omega^n_{\alpha\beta}$. As we shall see, the rank-three
tensor $Q^{(2)}$ captures higher geometric data beyond the quantum
metric and Berry curvature.

\subsection{Connected amplitudes}
\label{sec:recursion}

\begin{figure*}[t]
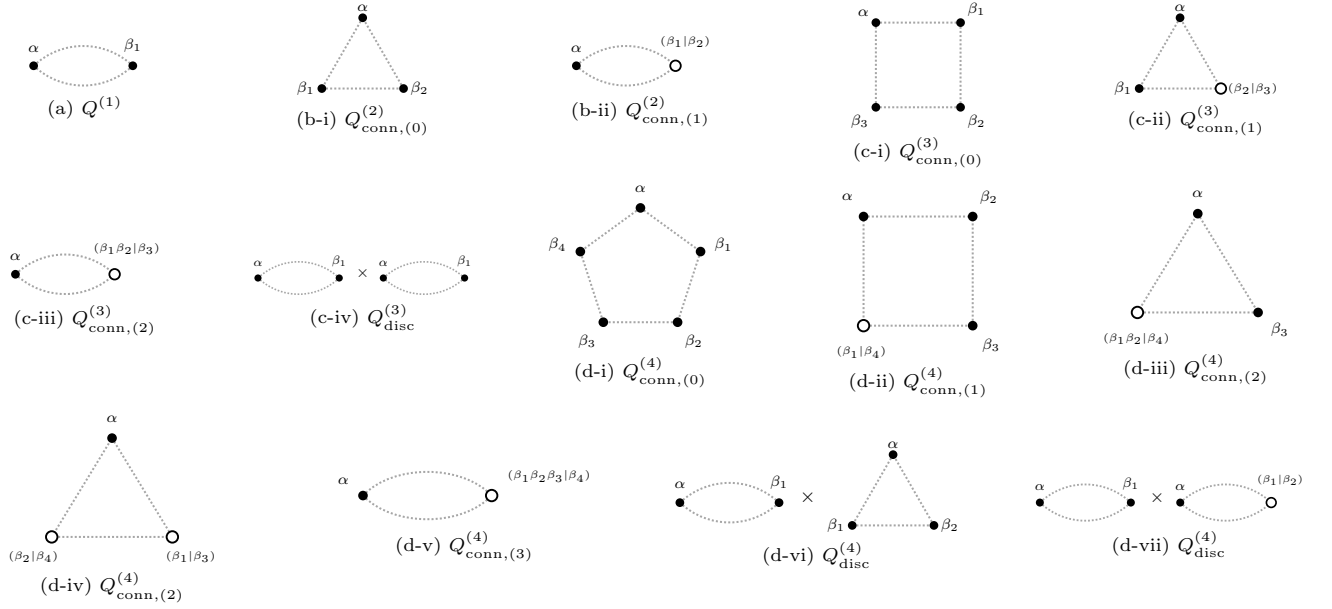

\centering

\begin{minipage}[t]{0.18\textwidth}\centering
\scalebox{0.80}{\SkelBubble{}}\\[0.02em]
{\scriptsize (a) $Q^{(1)}$}
\end{minipage}\hfill
\begin{minipage}[t]{0.18\textwidth}\centering
\scalebox{0.80}{\SkelTriangle{}}\\[0.02em]
{\scriptsize (b-i) $Q^{(2)}_{\conn,(0)}$}
\end{minipage}\hfill
\begin{minipage}[t]{0.18\textwidth}\centering
\scalebox{0.80}{\SkelCovBubbleOne}\\[0.02em]
{\scriptsize (b-ii) $Q^{(2)}_{\conn,(1)}$}
\end{minipage}\hfill
\begin{minipage}[t]{0.18\textwidth}\centering
\scalebox{0.80}{\SkelSquare{}}\\[0.02em]
{\scriptsize (c-i) $Q^{(3)}_{\conn,(0)}$}
\end{minipage}\hfill
\begin{minipage}[t]{0.18\textwidth}\centering
\scalebox{0.80}{\SkelCovTriangleOne}\\[0.02em]
{\scriptsize (c-ii) $Q^{(3)}_{\conn,(1)}$}
\end{minipage}

\vspace{0.35em}

\begin{minipage}[t]{0.18\textwidth}\centering
\scalebox{0.80}{\SkelCovBubbleTwo}\\[0.02em]
{\scriptsize (c-iii) $Q^{(3)}_{\conn,(2)}$}
\end{minipage}\hfill
\begin{minipage}[t]{0.18\textwidth}\centering
\resizebox{0.94\linewidth}{!}{\raisebox{-0.2em}{\SkelBubble{}}\;\(\times\)\;\raisebox{-0.2em}{\SkelBubble{}}}\\[0.02em]
{\scriptsize (c-iv) $Q^{(3)}_{\disc}$}
\end{minipage}\hfill
\begin{minipage}[t]{0.18\textwidth}\centering
\scalebox{0.80}{\SkelQfourConnZero}\\[0.02em]
{\scriptsize (d-i) $Q^{(4)}_{\conn,(0)}$}
\end{minipage}\hfill
\begin{minipage}[t]{0.18\textwidth}\centering
\scalebox{0.80}{\SkelQfourConnOne}\\[0.02em]
{\scriptsize (d-ii) $Q^{(4)}_{\conn,(1)}$}
\end{minipage}\hfill
\begin{minipage}[t]{0.18\textwidth}\centering
\scalebox{0.80}{\SkelQfourConnTwoA}\\[0.02em]
{\scriptsize (d-iii) $Q^{(4)}_{\conn,(2)}$}
\end{minipage}

\vspace{0.35em}

\begin{minipage}[t]{0.22\textwidth}\centering
\scalebox{0.80}{\SkelQfourConnTwoB}\\[0.02em]
{\scriptsize (d-iv) $Q^{(4)}_{\conn,(2)}$}
\end{minipage}\hfill
\begin{minipage}[t]{0.22\textwidth}\centering
\scalebox{0.80}{\SkelQfourConnThree}\\[0.02em]
{\scriptsize (d-v) $Q^{(4)}_{\conn,(3)}$}
\end{minipage}\hfill
\begin{minipage}[t]{0.22\textwidth}\centering
\scalebox{0.80}{\raisebox{-0.2em}{\SkelBubble{}}\;\(\times\)\;\raisebox{-0.2em}{\SkelTriangle{}}}\\[0.02em]
{\scriptsize (d-vi) $Q^{(4)}_{\disc}$}
\end{minipage}\hfill
\begin{minipage}[t]{0.22\textwidth}\centering
\resizebox{0.94\linewidth}{!}{\raisebox{-0.2em}{\SkelBubble{}}\;\(\times\)\;\raisebox{-0.2em}{\SkelCovBubbleOne}}\\[0.02em]
{\scriptsize (d-vii) $Q^{(4)}_{\disc}$}
\end{minipage}

\caption{Bargmann expansion through fourth order. Gray dotted lines denote projective contractions (intermediate-band sums) at fixed $n$ and $\kk$; they carry no energy denominators. A black vertex labelled $\mu$ denotes $A_\mu$, while an open vertex labelled $(\rho_1\cdots\rho_r|\mu)$ denotes $D_{\rho_1}\cdots D_{\rho_r}A_\mu$. Panels (a), (b-i)--(b-ii), (c-i)--(c-iv), and (d-i)--(d-vii) collect the $Q^{(1)}$, $Q^{(2)}$, $Q^{(3)}$, and $Q^{(4)}$ families. Panels (d-iii) and (d-iv) show the two inequivalent derivative distributions contributing to $Q^{(4)}_{\conn,(2)}$: one doubly differentiated vertex and two singly differentiated vertices, respectively.}
\label{fig:bargmann-skeletons-main}
\end{figure*}

To isolate the irreducible geometric data, we separate the amplitudes into
connected pieces and return-through-band products. The full derivation of the
recursion below is given in Sec.~\ref{app:recursion} of the Supplemental Material. For compactness, write
the ordered set of derivative directions as
\[
\bm{\beta}\equiv(\beta_1,\ldots,\beta_N),
\qquad
\bm{\beta}'\equiv(\beta_2,\ldots,\beta_N).
\]
Define the open-chain amplitude
\begin{equation}
\Lambda^{(N),n}_{m;\bm{\beta}}
=
\langle u_m|
\del_{\beta_1}\cdots\del_{\beta_N}P_n
|u_n\rangle,
\qquad m\ne n .
\label{eq:LambdaDef}
\end{equation}
Since $\langle u_n|\del_\alpha P_n|u_n\rangle=0$, the local tensor may
be closed through the Berry connection
$A_{\alpha,nm}\equiv i\langle u_n|\del_\alpha u_m\rangle$ as
\begin{equation}
Q^{(N),n}_{\alpha;\bm{\beta}}
=
\sum_{m\ne n}
iA_{\alpha,nm}
\Lambda^{(N),n}_{m;\bm{\beta}}.
\label{eq:QNviaLambda}
\end{equation}
The full amplitude obeys
\begin{equation}
\begin{split}
\Lambda^{(N)}_{m;\bm{\beta}}
&=
D_{\beta_1}
\Lambda^{(N-1)}_{m;\bm{\beta}'}  \\
&\quad
-i\!\sum_{\substack{l\ne n\\ l\ne m}}
A_{\beta_1,ml}
\Lambda^{(N-1)}_{l;\bm{\beta}'}
+R^{(N)}_{m;\bm{\beta}},
\end{split}
\label{eq:fullRecMain}
\end{equation}
with base case
\begin{equation}
\Lambda^{(1)}_{m;\beta}=-iA_{\beta,mn}.
\label{eq:LambdaBase}
\end{equation}
For an object transforming as
$X_{mn}\to e^{i(\chi_n-\chi_m)}X_{mn}$, the covariant derivative is
\begin{equation}
D_\beta X_{mn}
\equiv
\del_\beta X_{mn}
+i(A_{\beta,nn}-A_{\beta,mm})X_{mn}.
\label{eq:covD}
\end{equation}
The residue $R^{(N)}$ collects paths that leave band $n$, return to $n$,
and then leave again. Such paths factorize into lower-rank data and hence do
not carry new irreducible geometry. The connected amplitude is defined by
removing this return-through-$n$ residue and obeys
\begin{equation}
\begin{split}
\Lambda^{(N)}_{\conn;m;\bm{\beta}}
&=
D_{\beta_1}
\Lambda^{(N-1)}_{\conn;m;\bm{\beta}'}  \\
&\quad
-i\!\sum_{\substack{l\ne n\\ l\ne m}}
A_{\beta_1,ml}
\Lambda^{(N-1)}_{\conn;l;\bm{\beta}'}.
\end{split}
\label{eq:connRec}
\end{equation}
The return sector vanishes through $N=2$ and first contributes at $N=3$. Using Eq.~\eqref{eq:connRec}, the local tensor separates as
\begin{equation}
Q^{(N),n}_{\alpha;\bm{\beta}}
=
Q^{(N),n}_{\conn,\alpha;\bm{\beta}}
+
Q^{(N),n}_{\disc,\alpha;\bm{\beta}}.
\label{eq:Qsplit}
\end{equation}
The disconnected part is polynomial in lower-order connected amplitudes.
At fixed $N$, the connected hierarchy is graded by the total number $r$ of
covariant derivatives,
\begin{align}
Q^{(N),n}_{\conn,\alpha;\bm{\beta}}
&=
\sum_{r=0}^{N-1}
Q^{(N),n}_{\conn,(r),\alpha;\bm{\beta}},
\nonumber\\
Q^{(N),n}_{\conn,(r),\alpha;\bm{\beta}}
&=
\sum_{m\ne n}
iA_{\alpha,nm}
\Lambda^{(N),n}_{\conn,(r);m;\bm{\beta}}.
\label{eq:Qgraded}
\end{align}
Figure~\ref{fig:bargmann-skeletons-main} displays the resulting connected
and return-through-band diagrams up to fifth rank. The bubble in
panel (a) is the rank-two quantum geometric tensor, while the triangle in
panel (b-i) is its first irreducible three-band extension. Open vertices in
panels (b-ii), (c-ii), and beyond denote covariant derivatives acting along
a band path; the product diagrams in panels (c-iv), (d-vi), and (d-vii) are the
factorized return-through-band sector rather than new connected geometry.

\section{Geometric perturbation theory}
\label{sec:geomPT}

The $Q^{(N)}$ tensors discussed in the previous section describe the geometry
of virtual interband paths. To turn these geometric paths into physical
perturbation theory corrections, we attach field strengths and energy
denominators. Spectral weighting of their virtual interband segments defines
\begin{equation}
\bar Q^{(N)}_{(r)}\big|_{p_1\cdots p_N},
\qquad
\tilde Q^{(N)}_{(r)}\big|_{p_1\cdots p_N}.
\label{eq:Qweight}
\end{equation}
The barred tensors use bare gaps
$\Delta_{nm}=\epsilon_n^{(0)}-\epsilon_m^{(0)}$, while the tilded tensors
use dressed gaps $\Dtilde_{nm}=\epsilon_n-\epsilon_m^{(0)}$. The pole
labels $(p_1,\ldots,p_N)$ record the powers of the corresponding
denominators; the default $p=1$ is omitted. Symmetrization $\cS$ over
perturbation indices and the appropriate Hermitian projection are implicit.
Explicit forms of the tilded and barred tensors, and the relation between
them, are given in Sec.~\ref{app:diagrammatics} of the Supplemental Material.

\subsection{Vertex-order corrections}

Take a Hamiltonian with a spectrum indexed by band $n$ and a
perturbation of the form
\begin{equation}
V_{mn}=\alpha^a A_{a,mn}\quad(m\ne n),\qquad V_{nn}=0,
\label{eq:Vform}
\end{equation}
for a real expansion parameter $\alpha^a$. We work locally in the
parallel-transport gauge along the perturbation direction,
$\alpha^aA_{a,nn}=0$ for every band, so that $V_{nn}=0$. In this gauge the perturbation
has no intraband matrix elements and only mixes band $n$ with other bands $m$.

Each vertex is a perturbation insertion and therefore an interband
transition. An energy correction is
represented by a closed virtual path beginning and ending in $n$. Because
$V_{nn}=0$, the first nonzero path has two vertices, $n\to m\to n$; higher
orders insert additional interband transitions before the return. Brillouin--Wigner perturbation theory, summarized in Sec.~\ref{app:diagrammatics} of the Supplemental
Material, gives the vertex-order energy correction
as a single dressed connected loop,
\begin{equation}
T_n^{[N+1]}
=
\tilde Q^{(N),n}_{(0)\,a_0\cdots a_N}
\alpha^{a_0}\cdots\alpha^{a_N}.
\label{eq:Tdef}
\end{equation}
The same construction gives the dressed Berry connection, which carries a
free connection index $c$. In the $r=0$ sector this index can occupy any
of the $N+1$ vertices of the derivative-free loop, producing the
coefficient $N+1$. The connected vertex-order result is
\begin{align}
A_{c,n}^{[N]}\big|_{\conn}
&=
\tilde Q^{(N),n}_{\conn,c}\alpha^N,
\nonumber\\
\tilde Q^{(N),n}_{\conn,c}
&\equiv
(N+1)\tilde Q^{(N),n}_{(0);c}
+
\sum_{r=1}^{N-1}\tilde Q^{(N),n}_{(r);c}.
\label{eq:tildeQconn}
\end{align}
The corresponding vertex-order energy and connection diagrams are shown in
Sec.~\ref{app:diagrammatics} of the Supplemental Material.

\subsection{Strict-order corrections}

\begin{figure*}[t]
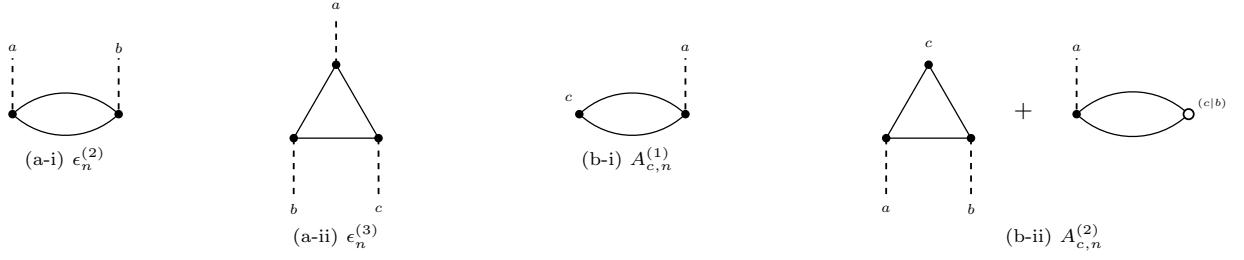

\centering
\begin{minipage}[t]{0.17\textwidth}
\centering
\scalebox{0.78}{\EnergyBubbleBare}\\[0.05em]
{\scriptsize (a-i) $\epsilon_n^{(2)}$}
\end{minipage}\hfill
\begin{minipage}[t]{0.17\textwidth}
\centering
\scalebox{0.78}{\EnergyTriangleBare}\\[0.05em]
{\scriptsize (a-ii) $\epsilon_n^{(3)}$}
\end{minipage}\hfill
\begin{minipage}[t]{0.20\textwidth}
\centering
\scalebox{0.78}{\ConnBubbleBare}\\[0.05em]
{\scriptsize (b-i) $A^{(1)}_{c,n}$}
\end{minipage}\hfill
\begin{minipage}[t]{0.37\textwidth}
\centering
\scalebox{0.78}{\ConnTriangleBare}
\quad \(+\) \quad
\scalebox{0.78}{\ConnCovBubbleBare}\\[0.05em]
{\scriptsize (b-ii) $A^{(2)}_{c,n}$}
\end{minipage}

\caption{Strict-order diagrams needed for the second- and third-order charge conductivity.
These are the local geometric paths of Fig.~\ref{fig:bargmann-skeletons-main}
after attaching external perturbation legs and weighting their interband
segments by bare resolvents. Panels (a-i) and (a-ii) are the first two nonzero energy corrections, while panels
(b-i) and (b-ii) are the connection corrections through second order. Thin
loop lines carry intermediate-band sums weighted by bare gaps
$1/\Delta_{nm}$. Dashed legs label external
perturbation weights $\alpha^a$, and the free connection index $c$ is
written.}
\label{fig:strict-order-main}
\end{figure*}

Strict order in $\alpha$ is obtained first by reexpanding the dressed
energy gaps in bare gaps. Expanding the normalization denominator of the
Berry connection supplies a second source of disconnected products. The
derivation and diagrammatic form of both operations are given in
Sec.~\ref{app:diagrammatics} of the Supplemental Material. Their leading structure is
\begin{align}
\epsilon_n^{(N+1)}
&=
\bar Q^{(N),n}_{(0)}\alpha^{N+1}
\nonumber\\
&\quad-
\sum_{K_1K_2}^{\prime}
\cS\!\left[
\bar Q^{(K_1),n}_{(0)}
\times
\bar Q^{(K_2),n}_{(0)}\big|_{2}
\right]
\alpha^{N+1}
+\cdots,
\label{eq:strictE}
\\[4pt]
A_{c,n}^{(N)}
&=
\bar Q^{(N),n}_{\conn,c}\alpha^N
\nonumber\\
&\quad-
\sum_{K_1K_2}^{\prime}
\cS\!\left[
\bar Q^{(K_1),n}_{\conn,c}\big|_{2}
\bar Q^{(K_2),n}_{(0)}
\right.
\nonumber\\[-2pt]
&\hspace{4.4em}\left.
+
\bar Q^{(K_1),n}_{\conn,c}
\bar Q^{(K_2),n}_{(0)}\big|_{2}
\right]
\alpha^N
+\cdots .
\label{eq:strictA}
\end{align}
Here $\sum_{K_1K_2}^{\prime}$ denotes the constrained sum with
$K_1+K_2=N-1$ and $K_1,K_2\ge1$. The omitted terms are higher products
of lower-order connected data generated by the same reexpansions. For the orders needed in the next section, the constrained sums are empty, so
strict order is obtained simply by replacing the dressed gaps in the
connected diagrams by bare gaps. The resulting diagrams are
shown in Fig.~\ref{fig:strict-order-main}.

\section{Nonlinear transport}
\label{sec:nonlinearTransport}

We now specialize the general perturbation to a uniform electric field,
$\alpha^a=-eE^a$, and quote the nonlinear charge conductivities through
third order. The
semiclassical wave-packet and Boltzmann derivation is given in
Sec.~\ref{app:transportDerivation} of the Supplemental Material, where the current is organized at
general order in the electric field before being evaluated explicitly at
second and third order. In the response formulas below,
$\epsilon_n$, $\Omega^n_{ab}$, and $f_n(\kk)$ denote the unperturbed band
energy, Berry curvature, and equilibrium Fermi--Dirac distribution,
respectively. We use the
shorthand
\begin{equation}
\langle X\rangle_f
\equiv
\sum_n
\int_{\BZ}\frac{d^d k}{(2\pi)^d}\,
f_n(\kk)\,X_n(\kk).
\label{eq:FermiSeaBracket}
\end{equation}
The formulas below hold in arbitrary dimension $d$, with the tensor indices
ranging over the corresponding momentum directions. Specializing the geometric tensors of Secs.~\ref{sec:local}
and~\ref{sec:geomPT} gives
\begin{align}
\bar Q^{(1),n}_{(0)\,ab}
&=
\tfrac{1}{2}\!\sum_{m\neq n}\!
\frac{A_{a,nm}A_{b,mn}+A_{b,nm}A_{a,mn}}{\Delta_{nm}},
\label{eq:Q1zero}\\[2pt]
\bar Q^{(2),n}_{(0)\,abc}
&=
\frac{1}{6}\sum_{\pi\in S_3}
\sum_{\substack{m_1,m_2\neq n\\m_1\neq m_2}}
\nonumber\\[-2pt]
&\quad\times
\frac{
A_{\pi(a),nm_1}
A_{\pi(b),m_1m_2}
A_{\pi(c),m_2n}}
{\Delta_{nm_1}\Delta_{nm_2}},
\label{eq:Q2zero}\\
\bar Q^{(2),n}_{(1);c\,ab}\big|_2
&=
-\tfrac{1}{2}\!\sum_{m\neq n}\!
\frac{\Imm\!\bigl[A_{a,nm}(D_c A_b)_{mn}\bigr]
+(a\!\leftrightarrow\!b)}
{\Delta_{nm}^{\,2}}.
\label{eq:Q2one}
\end{align}
Because the return sector vanishes through $N=2$, the second-order tensor is
fully connected. The combination entering the second-order
Berry-connection correction is
\begin{equation}
\bar Q^{(2),n}_{\conn,c;ab}
=
3\,\bar Q^{(2),n}_{(0)\,abc}
+
\bar Q^{(2),n}_{(1);c\,ab}\big|_2.
\label{eq:Q2conn}
\end{equation}
At $N=1$, the unweighted tensor is
$Q^{(1),n}_{a;b}=g^n_{ab}-\tfrac{i}{2}\Omega^n_{ab}$. The barred tensor
$\bar Q^{(1),n}_{(0)ab}$ used here is its gap-weighted Hermitian,
field-symmetric projection and is therefore the real band-normalized metric,
not the full complex quantum geometric tensor. By contrast, the explicit bare
curvatures in the BCD and BCQ channels enter directly through the equilibrium
anomalous velocity and carry no perturbative spectral weighting.

\subsection{Second-order response}

The second-order current is defined by
\begin{equation}
j_c^{(2)}=
\sigma^{ab;c}E_aE_b,
\label{eq:sigma2def}
\end{equation}
and decomposes according to its powers of the relaxation time as
\begin{equation}
\sigma^{ab;c}
=
\sigma_{D}^{ab;c}
+\sigma_{\rm BCD}^{ab;c}
+\sigma_{\rm QMD}^{ab;c}.
\label{eq:sigma2split}
\end{equation}
The nonlinear Drude channel is
\begin{equation}
\sigma_{D}^{ab;c}
=-
\frac{e^3\tau^2}{\hbar^3}
\left\langle
\del_a\del_b\del_c\epsilon_n
\right\rangle_f,
\label{eq:sigma2Drude}
\end{equation}
the Berry-curvature-dipole channel is~\cite{SodemannFuPRL2015}
\begin{equation}
\sigma_{\rm BCD}^{ab;c}
=
-
\frac{e^3\tau}{2\hbar^2}
\left\langle
\del_a\Omega^n_{bc}
+\del_b\Omega^n_{ac}
\right\rangle_f,
\label{eq:sigma2BCD}
\end{equation}
and the intrinsic quantum-metric-dipole channel is~\cite{LiuYuPRL2021,WangPRL2022}

\begin{equation}
\begin{split}
\sigma_{\rm QMD}^{ab;c}
&=-
\frac{e^3}{\hbar}
\Big\langle
\del_a\bar Q^{(1),n}_{(0)\,bc}
+
\del_b\bar Q^{(1),n}_{(0)\,ac}
\\[-2pt]
&\hspace{8.0em}
-
\del_c\bar Q^{(1),n}_{(0)\,ab}
\Big\rangle_f.
\end{split}
\label{eq:sigma2QMD}
\end{equation}
These are, respectively, the $\tau^2$, $\tau$, and $\tau^0$ sectors. No
$Q^{(2)}$ or higher tensor appears: at the level of the local geometric
hierarchy, the response is completely determined by rank-two data and can be
captured in a two-band model.

\subsection{Third-order response}

The third-order current is defined by
\begin{equation}
j_d^{(3)}=
\sigma^{abc;d}E_aE_bE_c,
\label{eq:sigma3def}
\end{equation}
and the result decomposes into five channels,
\begin{equation}
\sigma^{abc;d}
=
\sigma_D^{abc;d}
+
\sigma_{\rm BCQ}^{abc;d}
+
\sigma_{\rm QMQ}^{abc;d}
+
\sigma_{\cW}^{abc;d}
+
\sigma_{\cL}^{abc;d}.
\label{eq:sigma3split}
\end{equation}
The Drude channel is
\begin{equation}
\sigma_D^{abc;d}
=
\frac{e^4\tau^3}{\hbar^4}
\left\langle
\del_a\del_b\del_c\del_d\epsilon_n
\right\rangle_f .
\label{eq:sigmaD}
\end{equation}
The Berry curvature quadrupole channel
is~\cite{FangCanoPRL2024}
\begin{equation}
\begin{split}
\sigma_{\rm BCQ}^{abc;d}
&=
\frac{e^4\tau^2}{3\hbar^3}
\Big\langle
\del_a\del_b\Omega^n_{cd}
+
\del_b\del_c\Omega^n_{ad}
\\[-2pt]
&\hspace{7.4em}
+
\del_a\del_c\Omega^n_{bd}
\Big\rangle_f .
\end{split}
\label{eq:sigmaBCQ}
\end{equation}
The quantum metric quadrupole channel
is~\cite{FangCanoPRL2024,LiuBCPPRB2022}
\begin{equation}
\begin{split}
\sigma_{\rm QMQ}^{abc;d}
&=
\frac{e^4\tau}{3\hbar^2}
\Big\langle
\sum_{\mathrm{cyc}(a,b,c)}
\Big[
2\del_a\del_b\bar Q^{(1),n}_{(0)\,cd}
\\[-2pt]
&\hspace{8.0em}
-\del_a\del_d\bar Q^{(1),n}_{(0)\,bc}
\Big]
\Big\rangle_f .
\end{split}
\label{eq:sigmaQMQ}
\end{equation}
The first intrinsic rank-three channel is the curvature correction from
the second-order polarizability,
\begin{equation}
\sigma_{\cW}^{abc;d}
=
\frac{e^4}{3\hbar}
\left\langle
\cW^{(2),n}_{ad;bc}
+
\cW^{(2),n}_{bd;ac}
+
\cW^{(2),n}_{cd;ab}
\right\rangle_f ,
\label{eq:sigmaW}
\end{equation}
where
\begin{equation}
\cW^{(2),n}_{cd;ab}
\equiv
\del_c\bar Q^{(2),n}_{\conn,d;ab}
-
\del_d\bar Q^{(2),n}_{\conn,c;ab}.
\label{eq:WTwo}
\end{equation}
The second intrinsic rank-three channel is the gradient of the
three-band energy loop,
\begin{equation}
\sigma_{\cL}^{abc;d}
=
\frac{e^4}{\hbar}
\left\langle
\del_d\bar Q^{(2),n}_{(0)\,abc}
\right\rangle_f .
\label{eq:sigmaL}
\end{equation}
Thus, the $\cL$ channel is a total derivative controlled by the Fermi surface and vanishes for a completely filled band, for which $f_n$ is constant. The two $\tau^0$ terms are fully geometric rank-three channels. In a strict
two-band model the closed three-band loop of Eq.~\eqref{eq:Q2zero}
vanishes, so $\sigma_{\cL}^{abc;d}=0$. The remaining channel
$\sigma_{\cW}^{abc;d}$ reduces to the two-band contribution proportional
to $g\,\Omega/\Delta$ of~\cite{FangCanoPRL2024}. Thus the $\cL$
channel is irreducibly three-band, whereas the $\cW$ channel survives in a
two-band model and admits multiband corrections.

It is worthwhile to note that the two intrinsic channels have the following geometric interpretation. The dressed
projector
$\Pi_n=|\psi_n\rangle\langle\psi_n|/\langle\psi_n|\psi_n\rangle$, where $|\psi_n\rangle$ denotes the field-dressed Bloch state,
defines a field-dependent rotation of the local band frame. As $\kk$ is
driven by the electric field, virtual interband motion carries the wave
packet into a comoving frame in momentum space. At first order, the
field-induced correction to the Berry connection gives the leading
positional shift of the wave-packet center; order by order, the higher
connection corrections encode the corresponding higher-order shifts
\cite{GaoYangNiuPRL2014,LiuBCPPRB2022}. The connected tensors describe
the irreducible multiband data of this frame change. Both intrinsic channels are evaluated with the
equilibrium distribution, without a nonequilibrium population
correction. They therefore remain finite in
the formal $\tau\to0$ limit, which is the sense in which they are fully
geometric. 

\subsection{Symmetry transformations of the local geometry}
\label{sec:Qsymmetry}

The transformation law of the local tensors follows directly from their
projector definition. For a unitary crystalline symmetry $g$ represented in
momentum space by $R_g$, symmetry of the Hamiltonian implies
\begin{equation}
P_{gn}(R_g\kk)=U_gP_n(\kk)U_g^{-1}.
\label{eq:unitaryProjectorSymmetry}
\end{equation}
Here $gn$ denotes the symmetry image of band $n$. For the nondegenerate bands
considered here, $U_g|u_{n\kk}\rangle$ is, up to phase, the unique eigenstate
at $R_g\kk$ with energy $\epsilon_n(\kk)$, so no band mixing occurs. 

Applying the chain rule to Eq.~\eqref{eq:QNdef} and using cyclicity of the
trace gives
\begin{equation}
\begin{split}
Q^{(N),gn}_{a;b_1\cdots b_N}(R_g\kk)
&=(R_g)_{a\alpha}(R_g)_{b_1\beta_1}\cdots
(R_g)_{b_N\beta_N}
\\[-2pt]
&\quad\times
Q^{(N),n}_{\alpha;\beta_1\cdots\beta_N}(\kk).
\end{split}
\label{eq:QunitaryTransform}
\end{equation}
 For an
antiunitary symmetry $\Theta$ with momentum action $R_\Theta$, the same
argument applies, but the trace is complex conjugated:
\begin{equation}
\begin{split}
Q^{(N),\Theta n}_{a;b_1\cdots b_N}(R_\Theta\kk)
&=(R_\Theta)_{a\alpha}(R_\Theta)_{b_1\beta_1}\cdots
(R_\Theta)_{b_N\beta_N}
\\[-2pt]
&\quad\times
\left[Q^{(N),n}_{\alpha;\beta_1\cdots\beta_N}(\kk)\right]^*.
\end{split}
\label{eq:QantiunitaryTransform}
\end{equation}
For inversion $\mathcal P$, $R_{\mathcal P}=-I$ and the symmetry constraint
is
\begin{equation}
Q^{(N),\mathcal P n}(-\kk)
=(-1)^{N+1}Q^{(N),n}(\kk).
\label{eq:Qinversion}
\end{equation}
For time reversal $\mathcal T$,
\begin{equation}
Q^{(N),\mathcal T n}(-\kk)
=(-1)^{N+1}\left[Q^{(N),n}(\kk)\right]^*.
\label{eq:QtimeReversal}
\end{equation}
Consequently, under inversion both the real and imaginary parts have parity
$(-1)^{N+1}$, whereas under time reversal
\begin{equation}
\Ree Q^{(N)}:\;(-1)^{N+1},
\qquad
\Imm Q^{(N)}:\;(-1)^N.
\label{eq:QrealImagParity}
\end{equation}
For $N=1$, this reproduces the familiar time-reversal-even quantum metric
and time-reversal-odd Berry curvature; higher $N$ alternate according to
Eq.~\eqref{eq:QrealImagParity}. These local constraints feed directly into nonlinear transport. An
$r$th-order charge conductivity has $r+1$ indices, so inversion
requires $\sigma^{(r)}=(-1)^{r+1}\sigma^{(r)}$. Every second-order
conductivity therefore vanishes in an inversion-symmetric crystal, while a
third-order tensor is inversion allowed and becomes the leading nonlinear
charge response. At $N=2$, both components of the local tensor are inversion
odd, but $\cL$ contains a gradient and $\cW$ a curl, so both intrinsic
third-order channels are inversion even. They are therefore allowed in
centrosymmetric crystals, subject to further constraints from the full
magnetic point group and possible band-structure cancellations. Finally, since
$\Imm Q^{(2)}$ is time-reversal even, the combinations entering $\cL$ and $\cW$ make the integrand time-reversal odd, and the contributions at $\kk$
and $-\kk$ cancel. Consequently, time-reversal symmetry must be broken for either intrinsic channel to be nonzero.

\section{Conclusion and outlook}
\label{sec:conclusion}

We have shown that a general off-diagonal perturbation of Bloch bands is
organized by the local geometry of the three-point Bargmann invariant.
The hierarchy $Q^{(N)}$, separated into connected
amplitudes and return-through-band products, generates the dressed
dispersion and Berry connection. For a uniform electric field, the
third-order conductivity contains the Drude, Berry curvature quadrupole,
and quantum metric quadrupole channels together with two intrinsic
rank-three channels, $\sigma_{\cW}$ and $\sigma_{\cL}$; the latter is
irreducibly three-band. This framework is especially natural when symmetry suppresses lower-order
geometric response. In $d$-wave altermagnets, rotation, time-reversal, and
inversion symmetries can make third-order transport the leading geometric
probe of the spin-split multiband structure
\cite{SmejkalLandscapePRX2022,FangCanoPRL2024}. The intrinsic
channels identified here are therefore natural experimental targets. 

\begin{acknowledgments}
Work by S.F. and T.L. is supported by AFOSR under MURI Grant No. FA9550-25-1-0262.
Work by E.M. is supported by the Department of Energy under Grant No. DE-FG02-84ER45118.
\end{acknowledgments}

\bibliographystyle{apsrev4-2}
\bibliography{refs}

\begin{thebibliography}{24}%
\makeatletter
\providecommand \@ifxundefined [1]{%
 \@ifx{#1\undefined}
}%
\providecommand \@ifnum [1]{%
 \ifnum #1\expandafter \@firstoftwo
 \else \expandafter \@secondoftwo
 \fi
}%
\providecommand \@ifx [1]{%
 \ifx #1\expandafter \@firstoftwo
 \else \expandafter \@secondoftwo
 \fi
}%
\providecommand \natexlab [1]{#1}%
\providecommand \enquote  [1]{``#1''}%
\providecommand \bibnamefont  [1]{#1}%
\providecommand \bibfnamefont [1]{#1}%
\providecommand \citenamefont [1]{#1}%
\providecommand \href@noop [0]{\@secondoftwo}%
\providecommand \href [0]{\begingroup \@sanitize@url \@href}%
\providecommand \@href[1]{\@@startlink{#1}\@@href}%
\providecommand \@@href[1]{\endgroup#1\@@endlink}%
\providecommand \@sanitize@url [0]{\catcode `\\12\catcode `\$12\catcode `\&12\catcode `\#12\catcode `\^12\catcode `\_12\catcode `\%12\relax}%
\providecommand \@@startlink[1]{}%
\providecommand \@@endlink[0]{}%
\providecommand \url  [0]{\begingroup\@sanitize@url \@url }%
\providecommand \@url [1]{\endgroup\@href {#1}{\urlprefix }}%
\providecommand \urlprefix  [0]{URL }%
\providecommand \Eprint [0]{\href }%
\providecommand \doibase [0]{https://doi.org/}%
\providecommand \selectlanguage [0]{\@gobble}%
\providecommand \bibinfo  [0]{\@secondoftwo}%
\providecommand \bibfield  [0]{\@secondoftwo}%
\providecommand \translation [1]{[#1]}%
\providecommand \BibitemOpen [0]{}%
\providecommand \bibitemStop [0]{}%
\providecommand \bibitemNoStop [0]{.\EOS\space}%
\providecommand \EOS [0]{\spacefactor3000\relax}%
\providecommand \BibitemShut  [1]{\csname bibitem#1\endcsname}%
\let\auto@bib@innerbib\@empty
\bibitem [{\citenamefont {Provost}\ and\ \citenamefont {Vall{\'e}e}(1980)}]{ProvostVallee1980}%
  \BibitemOpen
  \bibfield  {author} {\bibinfo {author} {\bibfnamefont {J.~P.}\ \bibnamefont {Provost}}\ and\ \bibinfo {author} {\bibfnamefont {G.}~\bibnamefont {Vall{\'e}e}},\ }\href {https://doi.org/10.1007/BF02193559} {\bibfield  {journal} {\bibinfo  {journal} {Commun. Math. Phys.}\ }\textbf {\bibinfo {volume} {76}},\ \bibinfo {pages} {289} (\bibinfo {year} {1980})}\BibitemShut {NoStop}%
\bibitem [{\citenamefont {Berry}(1984)}]{Berry1984}%
  \BibitemOpen
  \bibfield  {author} {\bibinfo {author} {\bibfnamefont {M.~V.}\ \bibnamefont {Berry}},\ }\href@noop {} {\bibfield  {journal} {\bibinfo  {journal} {Proc. R. Soc. Lond. A}\ }\textbf {\bibinfo {volume} {392}},\ \bibinfo {pages} {45} (\bibinfo {year} {1984})}\BibitemShut {NoStop}%
\bibitem [{\citenamefont {Simon}(1983)}]{Simon1983}%
  \BibitemOpen
  \bibfield  {author} {\bibinfo {author} {\bibfnamefont {B.}~\bibnamefont {Simon}},\ }\href@noop {} {\bibfield  {journal} {\bibinfo  {journal} {Phys. Rev. Lett.}\ }\textbf {\bibinfo {volume} {51}},\ \bibinfo {pages} {2167} (\bibinfo {year} {1983})}\BibitemShut {NoStop}%
\bibitem [{\citenamefont {Wilczek}\ and\ \citenamefont {Zee}(1984)}]{WilczekZeePRL1984}%
  \BibitemOpen
  \bibfield  {author} {\bibinfo {author} {\bibfnamefont {F.}~\bibnamefont {Wilczek}}\ and\ \bibinfo {author} {\bibfnamefont {A.}~\bibnamefont {Zee}},\ }\href@noop {} {\bibfield  {journal} {\bibinfo  {journal} {Phys. Rev. Lett.}\ }\textbf {\bibinfo {volume} {52}},\ \bibinfo {pages} {2111} (\bibinfo {year} {1984})}\BibitemShut {NoStop}%
\bibitem [{\citenamefont {Xiao}\ \emph {et~al.}(2010)\citenamefont {Xiao}, \citenamefont {Chang},\ and\ \citenamefont {Niu}}]{XiaoChangNiuRMP2010}%
  \BibitemOpen
  \bibfield  {author} {\bibinfo {author} {\bibfnamefont {D.}~\bibnamefont {Xiao}}, \bibinfo {author} {\bibfnamefont {M.-C.}\ \bibnamefont {Chang}},\ and\ \bibinfo {author} {\bibfnamefont {Q.}~\bibnamefont {Niu}},\ }\href {https://doi.org/10.1103/RevModPhys.82.1959} {\bibfield  {journal} {\bibinfo  {journal} {Rev. Mod. Phys.}\ }\textbf {\bibinfo {volume} {82}},\ \bibinfo {pages} {1959} (\bibinfo {year} {2010})}\BibitemShut {NoStop}%
\bibitem [{\citenamefont {Souza}\ \emph {et~al.}(2000)\citenamefont {Souza}, \citenamefont {Wilkens},\ and\ \citenamefont {Martin}}]{SouzaWilkensMartinPRB2000}%
  \BibitemOpen
  \bibfield  {author} {\bibinfo {author} {\bibfnamefont {I.}~\bibnamefont {Souza}}, \bibinfo {author} {\bibfnamefont {T.}~\bibnamefont {Wilkens}},\ and\ \bibinfo {author} {\bibfnamefont {R.~M.}\ \bibnamefont {Martin}},\ }\href {https://doi.org/10.1103/PhysRevB.62.1666} {\bibfield  {journal} {\bibinfo  {journal} {Phys. Rev. B}\ }\textbf {\bibinfo {volume} {62}},\ \bibinfo {pages} {1666} (\bibinfo {year} {2000})}\BibitemShut {NoStop}%
\bibitem [{\citenamefont {Peotta}\ and\ \citenamefont {T{\"o}rm{\"a}}(2015)}]{PeottaTormaNatCommun2015}%
  \BibitemOpen
  \bibfield  {author} {\bibinfo {author} {\bibfnamefont {S.}~\bibnamefont {Peotta}}\ and\ \bibinfo {author} {\bibfnamefont {P.}~\bibnamefont {T{\"o}rm{\"a}}},\ }\href@noop {} {\bibfield  {journal} {\bibinfo  {journal} {Nat. Commun.}\ }\textbf {\bibinfo {volume} {6}},\ \bibinfo {pages} {8944} (\bibinfo {year} {2015})}\BibitemShut {NoStop}%
\bibitem [{\citenamefont {Nagaosa}\ \emph {et~al.}(2010)\citenamefont {Nagaosa}, \citenamefont {Sinova}, \citenamefont {Onoda}, \citenamefont {MacDonald},\ and\ \citenamefont {Ong}}]{NagaosaAHE_RMP2010}%
  \BibitemOpen
  \bibfield  {author} {\bibinfo {author} {\bibfnamefont {N.}~\bibnamefont {Nagaosa}}, \bibinfo {author} {\bibfnamefont {J.}~\bibnamefont {Sinova}}, \bibinfo {author} {\bibfnamefont {S.}~\bibnamefont {Onoda}}, \bibinfo {author} {\bibfnamefont {A.~H.}\ \bibnamefont {MacDonald}},\ and\ \bibinfo {author} {\bibfnamefont {N.~P.}\ \bibnamefont {Ong}},\ }\href@noop {} {\bibfield  {journal} {\bibinfo  {journal} {Rev. Mod. Phys.}\ }\textbf {\bibinfo {volume} {82}},\ \bibinfo {pages} {1539} (\bibinfo {year} {2010})}\BibitemShut {NoStop}%
\bibitem [{\citenamefont {Thouless}\ \emph {et~al.}(1982)\citenamefont {Thouless}, \citenamefont {Kohmoto}, \citenamefont {Nightingale},\ and\ \citenamefont {den Nijs}}]{TKNNPRL1982}%
  \BibitemOpen
  \bibfield  {author} {\bibinfo {author} {\bibfnamefont {D.~J.}\ \bibnamefont {Thouless}}, \bibinfo {author} {\bibfnamefont {M.}~\bibnamefont {Kohmoto}}, \bibinfo {author} {\bibfnamefont {M.~P.}\ \bibnamefont {Nightingale}},\ and\ \bibinfo {author} {\bibfnamefont {M.}~\bibnamefont {den Nijs}},\ }\href {https://doi.org/10.1103/PhysRevLett.49.405} {\bibfield  {journal} {\bibinfo  {journal} {Phys. Rev. Lett.}\ }\textbf {\bibinfo {volume} {49}},\ \bibinfo {pages} {405} (\bibinfo {year} {1982})}\BibitemShut {NoStop}%
\bibitem [{\citenamefont {Kohmoto}(1985)}]{Kohmoto1985}%
  \BibitemOpen
  \bibfield  {author} {\bibinfo {author} {\bibfnamefont {M.}~\bibnamefont {Kohmoto}},\ }\href@noop {} {\bibfield  {journal} {\bibinfo  {journal} {Ann. Phys. (N.Y.)}\ }\textbf {\bibinfo {volume} {160}},\ \bibinfo {pages} {343} (\bibinfo {year} {1985})}\BibitemShut {NoStop}%
\bibitem [{\citenamefont {Niu}\ \emph {et~al.}(1985)\citenamefont {Niu}, \citenamefont {Thouless},\ and\ \citenamefont {Wu}}]{NiuThoulessWuPRB1985}%
  \BibitemOpen
  \bibfield  {author} {\bibinfo {author} {\bibfnamefont {Q.}~\bibnamefont {Niu}}, \bibinfo {author} {\bibfnamefont {D.~J.}\ \bibnamefont {Thouless}},\ and\ \bibinfo {author} {\bibfnamefont {Y.-S.}\ \bibnamefont {Wu}},\ }\href {https://doi.org/10.1103/PhysRevB.31.3372} {\bibfield  {journal} {\bibinfo  {journal} {Phys. Rev. B}\ }\textbf {\bibinfo {volume} {31}},\ \bibinfo {pages} {3372} (\bibinfo {year} {1985})}\BibitemShut {NoStop}%
\bibitem [{\citenamefont {Avron}\ \emph {et~al.}(1983)\citenamefont {Avron}, \citenamefont {Seiler},\ and\ \citenamefont {Simon}}]{Avron1983}%
  \BibitemOpen
  \bibfield  {author} {\bibinfo {author} {\bibfnamefont {J.~E.}\ \bibnamefont {Avron}}, \bibinfo {author} {\bibfnamefont {R.}~\bibnamefont {Seiler}},\ and\ \bibinfo {author} {\bibfnamefont {B.}~\bibnamefont {Simon}},\ }\href@noop {} {\bibfield  {journal} {\bibinfo  {journal} {Phys. Rev. Lett.}\ }\textbf {\bibinfo {volume} {51}},\ \bibinfo {pages} {51} (\bibinfo {year} {1983})}\BibitemShut {NoStop}%
\bibitem [{\citenamefont {Resta}\ and\ \citenamefont {Sorella}(1999)}]{RestaSorella1999}%
  \BibitemOpen
  \bibfield  {author} {\bibinfo {author} {\bibfnamefont {R.}~\bibnamefont {Resta}}\ and\ \bibinfo {author} {\bibfnamefont {S.}~\bibnamefont {Sorella}},\ }\href@noop {} {\bibfield  {journal} {\bibinfo  {journal} {Phys. Rev. Lett.}\ }\textbf {\bibinfo {volume} {82}},\ \bibinfo {pages} {370} (\bibinfo {year} {1999})}\BibitemShut {NoStop}%
\bibitem [{\citenamefont {Marzari}\ and\ \citenamefont {Vanderbilt}(1997)}]{MarzariVanderbiltPRB1997}%
  \BibitemOpen
  \bibfield  {author} {\bibinfo {author} {\bibfnamefont {N.}~\bibnamefont {Marzari}}\ and\ \bibinfo {author} {\bibfnamefont {D.}~\bibnamefont {Vanderbilt}},\ }\href@noop {} {\bibfield  {journal} {\bibinfo  {journal} {Phys. Rev. B}\ }\textbf {\bibinfo {volume} {56}},\ \bibinfo {pages} {12847} (\bibinfo {year} {1997})}\BibitemShut {NoStop}%
\bibitem [{\citenamefont {Marzari}\ \emph {et~al.}(2012)\citenamefont {Marzari}, \citenamefont {Mostofi}, \citenamefont {Yates}, \citenamefont {Souza},\ and\ \citenamefont {Vanderbilt}}]{MarzariRMP2012}%
  \BibitemOpen
  \bibfield  {author} {\bibinfo {author} {\bibfnamefont {N.}~\bibnamefont {Marzari}}, \bibinfo {author} {\bibfnamefont {A.~A.}\ \bibnamefont {Mostofi}}, \bibinfo {author} {\bibfnamefont {J.~R.}\ \bibnamefont {Yates}}, \bibinfo {author} {\bibfnamefont {I.}~\bibnamefont {Souza}},\ and\ \bibinfo {author} {\bibfnamefont {D.}~\bibnamefont {Vanderbilt}},\ }\href@noop {} {\bibfield  {journal} {\bibinfo  {journal} {Rev. Mod. Phys.}\ }\textbf {\bibinfo {volume} {84}},\ \bibinfo {pages} {1419} (\bibinfo {year} {2012})}\BibitemShut {NoStop}%
\bibitem [{\citenamefont {Bargmann}(1964)}]{Bargmann1964}%
  \BibitemOpen
  \bibfield  {author} {\bibinfo {author} {\bibfnamefont {V.}~\bibnamefont {Bargmann}},\ }\href@noop {} {\bibfield  {journal} {\bibinfo  {journal} {J. Math. Phys.}\ }\textbf {\bibinfo {volume} {5}},\ \bibinfo {pages} {862} (\bibinfo {year} {1964})}\BibitemShut {NoStop}%
\bibitem [{\citenamefont {Avdoshkin}\ and\ \citenamefont {Popov}(2023)}]{AvdoshkinPopovPRB2023}%
  \BibitemOpen
  \bibfield  {author} {\bibinfo {author} {\bibfnamefont {A.}~\bibnamefont {Avdoshkin}}\ and\ \bibinfo {author} {\bibfnamefont {F.~K.}\ \bibnamefont {Popov}},\ }\href {https://doi.org/10.1103/PhysRevB.107.245136} {\bibfield  {journal} {\bibinfo  {journal} {Phys. Rev. B}\ }\textbf {\bibinfo {volume} {107}},\ \bibinfo {pages} {245136} (\bibinfo {year} {2023})}\BibitemShut {NoStop}%
\bibitem [{\citenamefont {Fang}\ \emph {et~al.}(2024)\citenamefont {Fang}, \citenamefont {Cano},\ and\ \citenamefont {Ghorashi}}]{FangCanoPRL2024}%
  \BibitemOpen
  \bibfield  {author} {\bibinfo {author} {\bibfnamefont {Y.}~\bibnamefont {Fang}}, \bibinfo {author} {\bibfnamefont {J.}~\bibnamefont {Cano}},\ and\ \bibinfo {author} {\bibfnamefont {S.~A.~A.}\ \bibnamefont {Ghorashi}},\ }\href {https://doi.org/10.1103/PhysRevLett.133.106701} {\bibfield  {journal} {\bibinfo  {journal} {Phys. Rev. Lett.}\ }\textbf {\bibinfo {volume} {133}},\ \bibinfo {pages} {106701} (\bibinfo {year} {2024})}\BibitemShut {NoStop}%
\bibitem [{\citenamefont {Sodemann}\ and\ \citenamefont {Fu}(2015)}]{SodemannFuPRL2015}%
  \BibitemOpen
  \bibfield  {author} {\bibinfo {author} {\bibfnamefont {I.}~\bibnamefont {Sodemann}}\ and\ \bibinfo {author} {\bibfnamefont {L.}~\bibnamefont {Fu}},\ }\href@noop {} {\bibfield  {journal} {\bibinfo  {journal} {Phys. Rev. Lett.}\ }\textbf {\bibinfo {volume} {115}},\ \bibinfo {pages} {216806} (\bibinfo {year} {2015})}\BibitemShut {NoStop}%
\bibitem [{\citenamefont {Liu}\ \emph {et~al.}(2021)\citenamefont {Liu}, \citenamefont {Zhao}, \citenamefont {Huang}, \citenamefont {Wu}, \citenamefont {Sheng}, \citenamefont {Xiao},\ and\ \citenamefont {Yang}}]{LiuYuPRL2021}%
  \BibitemOpen
  \bibfield  {author} {\bibinfo {author} {\bibfnamefont {H.}~\bibnamefont {Liu}}, \bibinfo {author} {\bibfnamefont {J.}~\bibnamefont {Zhao}}, \bibinfo {author} {\bibfnamefont {Y.-X.}\ \bibnamefont {Huang}}, \bibinfo {author} {\bibfnamefont {W.}~\bibnamefont {Wu}}, \bibinfo {author} {\bibfnamefont {X.-L.}\ \bibnamefont {Sheng}}, \bibinfo {author} {\bibfnamefont {C.}~\bibnamefont {Xiao}},\ and\ \bibinfo {author} {\bibfnamefont {S.~A.}\ \bibnamefont {Yang}},\ }\href@noop {} {\bibfield  {journal} {\bibinfo  {journal} {Phys. Rev. Lett.}\ }\textbf {\bibinfo {volume} {127}},\ \bibinfo {pages} {277202} (\bibinfo {year} {2021})}\BibitemShut {NoStop}%
\bibitem [{\citenamefont {Wang}\ \emph {et~al.}(2021)\citenamefont {Wang}, \citenamefont {Gao},\ and\ \citenamefont {Xiao}}]{WangPRL2022}%
  \BibitemOpen
  \bibfield  {author} {\bibinfo {author} {\bibfnamefont {C.}~\bibnamefont {Wang}}, \bibinfo {author} {\bibfnamefont {Y.}~\bibnamefont {Gao}},\ and\ \bibinfo {author} {\bibfnamefont {D.}~\bibnamefont {Xiao}},\ }\href@noop {} {\bibfield  {journal} {\bibinfo  {journal} {Phys. Rev. Lett.}\ }\textbf {\bibinfo {volume} {127}},\ \bibinfo {pages} {277201} (\bibinfo {year} {2021})}\BibitemShut {NoStop}%
\bibitem [{\citenamefont {Liu}\ \emph {et~al.}(2022)\citenamefont {Liu}, \citenamefont {Zhao}, \citenamefont {Huang}, \citenamefont {Feng}, \citenamefont {Xiao}, \citenamefont {Wu}, \citenamefont {Lai}, \citenamefont {Gao},\ and\ \citenamefont {Yang}}]{LiuBCPPRB2022}%
  \BibitemOpen
  \bibfield  {author} {\bibinfo {author} {\bibfnamefont {H.}~\bibnamefont {Liu}}, \bibinfo {author} {\bibfnamefont {J.}~\bibnamefont {Zhao}}, \bibinfo {author} {\bibfnamefont {Y.-X.}\ \bibnamefont {Huang}}, \bibinfo {author} {\bibfnamefont {X.}~\bibnamefont {Feng}}, \bibinfo {author} {\bibfnamefont {C.}~\bibnamefont {Xiao}}, \bibinfo {author} {\bibfnamefont {W.}~\bibnamefont {Wu}}, \bibinfo {author} {\bibfnamefont {S.}~\bibnamefont {Lai}}, \bibinfo {author} {\bibfnamefont {W.-b.}\ \bibnamefont {Gao}},\ and\ \bibinfo {author} {\bibfnamefont {S.~A.}\ \bibnamefont {Yang}},\ }\href {https://doi.org/10.1103/PhysRevB.105.045118} {\bibfield  {journal} {\bibinfo  {journal} {Phys. Rev. B}\ }\textbf {\bibinfo {volume} {105}},\ \bibinfo {pages} {045118} (\bibinfo {year} {2022})}\BibitemShut {NoStop}%
\bibitem [{\citenamefont {Gao}\ \emph {et~al.}(2014)\citenamefont {Gao}, \citenamefont {Yang},\ and\ \citenamefont {Niu}}]{GaoYangNiuPRL2014}%
  \BibitemOpen
  \bibfield  {author} {\bibinfo {author} {\bibfnamefont {Y.}~\bibnamefont {Gao}}, \bibinfo {author} {\bibfnamefont {S.~A.}\ \bibnamefont {Yang}},\ and\ \bibinfo {author} {\bibfnamefont {Q.}~\bibnamefont {Niu}},\ }\href {https://doi.org/10.1103/PhysRevLett.112.166601} {\bibfield  {journal} {\bibinfo  {journal} {Phys. Rev. Lett.}\ }\textbf {\bibinfo {volume} {112}},\ \bibinfo {pages} {166601} (\bibinfo {year} {2014})}\BibitemShut {NoStop}%
\bibitem [{\citenamefont {{\v{S}}mejkal}\ \emph {et~al.}(2022)\citenamefont {{\v{S}}mejkal}, \citenamefont {Sinova},\ and\ \citenamefont {Jungwirth}}]{SmejkalLandscapePRX2022}%
  \BibitemOpen
  \bibfield  {author} {\bibinfo {author} {\bibfnamefont {L.}~\bibnamefont {{\v{S}}mejkal}}, \bibinfo {author} {\bibfnamefont {J.}~\bibnamefont {Sinova}},\ and\ \bibinfo {author} {\bibfnamefont {T.}~\bibnamefont {Jungwirth}},\ }\href {https://doi.org/10.1103/PhysRevX.12.040501} {\bibfield  {journal} {\bibinfo  {journal} {Phys. Rev. X}\ }\textbf {\bibinfo {volume} {12}},\ \bibinfo {pages} {040501} (\bibinfo {year} {2022})}\BibitemShut {NoStop}%
\end{thebibliography}%

\clearpage
\onecolumngrid

\begin{center}
\textbf{\large Supplemental Material for ``High Order Geometric Channels for Nonlinear Transport in Bloch Bands''}
\end{center}

\setcounter{section}{0}
\setcounter{equation}{0}
\setcounter{figure}{0}
\renewcommand{\thesection}{S\arabic{section}}
\renewcommand{\theequation}{S\arabic{equation}}
\renewcommand{\thefigure}{S\arabic{figure}}
\def\theHsection{S\arabic{section}}
\def\theHequation{S\arabic{equation}}
\def\theHfigure{S\arabic{figure}}

This Supplemental Material contains the algebraic derivation of the covariant
recursion behind the connected hierarchy (Sec.~\ref{app:recursion}), the
Brillouin--Wigner perturbative derivation and diagrammatic rules for the
vertex- and strict-order expansions (Sec.~\ref{app:diagrammatics}), and the
semiclassical derivation of nonlinear transport through third order
(Sec.~\ref{app:transportDerivation}).

\section{Algebraic derivation of the recursion}
\label{app:recursion}

For the ordered derivative directions
$\bm{\beta}\equiv(\beta_1,\ldots,\beta_N)$ and
$\bm{\beta}'\equiv(\beta_2,\ldots,\beta_N)$, the local tensor and open-chain
amplitude are
\[
\begin{aligned}
Q^{(N),n}_{\alpha;\bm{\beta}}
&=\tr\!\left[
P_n(\del_\alpha P_n)
(\del_{\beta_1}\cdots\del_{\beta_N}P_n)
\right],
\\[2pt]
\Lambda^{(N),n}_{m;\bm{\beta}}
&=\langle u_m|
\del_{\beta_1}\cdots\del_{\beta_N}P_n
|u_n\rangle,
\quad m\ne n.
\end{aligned}
\]
Inserting the band resolution of the identity between the two differentiated
projectors gives
\begin{equation}
\begin{split}
Q^{(N),n}_{\alpha;\bm{\beta}}
&=\sum_{m\ne n}
\langle u_n|\del_\alpha P_n|u_m\rangle
\langle u_m|\del_{\beta_1}\cdots\del_{\beta_N}P_n|u_n\rangle
\\
&=\sum_{m\ne n}iA_{\alpha,nm}
\Lambda^{(N),n}_{m;\bm{\beta}},
\end{split}
\label{eq:appQClosure}
\end{equation}
where the $m=n$ term vanishes because
$\langle u_n|\del_\alpha P_n|u_n\rangle=0$. Now write $X=\del_{\beta_2}\cdots\del_{\beta_N}P_n$. Differentiating
$\Lambda^{(N-1)}_{m;\bm{\beta}'}=\langle u_m|X|u_n\rangle$ gives
\begin{equation}
\begin{split}
\Lambda^{(N)}_{m;\bm{\beta}}
&=
\del_{\beta_1}\Lambda^{(N-1)}_{m;\bm{\beta}'}
-\langle\del_{\beta_1}u_m|X|u_n\rangle  \\
&\quad
-\langle u_m|X|\del_{\beta_1}u_n\rangle .
\end{split}
\label{eq:appLeibLambda}
\end{equation}
After inserting complete sets, the first two terms give
\begin{equation}
\begin{split}
&\del_{\beta_1}\Lambda^{(N-1)}_{m;\bm{\beta}'}
-\langle\del_{\beta_1}u_m|X|u_n\rangle  \\
&=
\del_{\beta_1}\Lambda^{(N-1)}_{m;\bm{\beta}'}
-iA_{\beta_1,mm}\Lambda^{(N-1)}_{m;\bm{\beta}'}  \\
&\quad
-i\!\sum_{\substack{l\ne n\\ l\ne m}}
A_{\beta_1,ml}\Lambda^{(N-1)}_{l;\bm{\beta}'}
-iA_{\beta_1,mn}\langle u_n|X|u_n\rangle ,
\end{split}
\label{eq:appFirstTwoTerms}
\end{equation}
while the last term gives
\begin{equation}
\begin{split}
-\langle u_m|X|\del_{\beta_1}u_n\rangle
&=
iA_{\beta_1,nn}\Lambda^{(N-1)}_{m;\bm{\beta}'}  \\
&\quad
+i\sum_{l\ne n}
\langle u_m|X|u_l\rangle A_{\beta_1,ln}.
\end{split}
\label{eq:appLastTerm}
\end{equation}
The diagonal pieces combine into the covariant derivative
\begin{equation}
D_\beta X_{mn}
=
\del_\beta X_{mn}
+i(A_{\beta,nn}-A_{\beta,mm})X_{mn},
\label{eq:appCovD}
\end{equation}
for an object transforming as
$X_{mn}\to e^{i(\chi_n-\chi_m)}X_{mn}$. Combining everything gives
\begin{equation}
\begin{split}
\Lambda^{(N)}_{m;\bm{\beta}}
&=
D_{\beta_1}\Lambda^{(N-1)}_{m;\bm{\beta}'}
-i\!\sum_{\substack{l\ne n\\ l\ne m}}
A_{\beta_1,ml}\Lambda^{(N-1)}_{l;\bm{\beta}'}  \\
&\quad
+R^{(N)}_{m;\bm{\beta}},
\end{split}
\label{eq:appFullLambdaRec}
\end{equation}
with
\begin{equation}
\begin{split}
R^{(N)}_{m;\bm{\beta}}
&=
-iA_{\beta_1,mn}\langle u_n|X|u_n\rangle  \\
&\quad
+i\sum_{l\ne n}
\langle u_m|X|u_l\rangle A_{\beta_1,ln}.
\end{split}
\label{eq:appRterm}
\end{equation}
For $N=2$, $X=\del_{\beta_2}P_n$ and both pieces in
Eq.~\eqref{eq:appRterm} vanish. Thus the return sector vanishes
through $N=2$. At $N=3$, however,
$X=\del_{\beta_2}\del_{\beta_3}P_n$, and both matrix elements
entering Eq.~\eqref{eq:appRterm} are nonzero:
\begin{align}
\langle u_n|X|u_n\rangle
&=
-\sum_{p\ne n}
\Big[
A_{\beta_2,np}A_{\beta_3,pn}
+
A_{\beta_3,np}A_{\beta_2,pn}
\Big],
\\
\langle u_m|X|u_l\rangle
&=
A_{\beta_3,mn}A_{\beta_2,nl}
+
A_{\beta_2,mn}A_{\beta_3,nl},
\qquad m,l\ne n.
\end{align}
Substitution into Eq.~\eqref{eq:appRterm} gives
\begin{align}
R^{(3)}_{m;\beta_1\beta_2\beta_3}
&=
iA_{\beta_1,mn}
\sum_{p\ne n}
\Big[
A_{\beta_2,np}A_{\beta_3,pn}
+
A_{\beta_3,np}A_{\beta_2,pn}
\Big]
\nonumber\\
&\quad+
i\sum_{l\ne n}
\Big[
A_{\beta_3,mn}A_{\beta_2,nl}
+
A_{\beta_2,mn}A_{\beta_3,nl}
\Big]
A_{\beta_1,ln}.
\end{align}
When the open chain is closed by $iA_{\alpha,nm}$ in
Eq.~\eqref{eq:appQClosure}, every term factorizes into two
closed paths through band $n$. For example,
\begin{equation}
iA_{\alpha,nm}\,
iA_{\beta_3,mn}A_{\beta_2,nl}A_{\beta_1,ln}
=
-
\bigl(A_{\alpha,nm}A_{\beta_3,mn}\bigr)
\bigl(A_{\beta_2,nl}A_{\beta_1,ln}\bigr).
\end{equation}
The two factors describe the separate returns
$n\to m\to n$ and $n\to l\to n$. Hence $R^{(3)}$ belongs to
the disconnected return-through-$n$ sector and contains no new
connected geometry.

\section{Brillouin--Wigner perturbation and diagrammatic rules}
\label{app:diagrammatics}

\subsection{Energy corrections}

We take $V_{mn}=\alpha^aA_{a,mn}$ for $m\ne n$ and work locally in the
parallel-transport gauge $\alpha^aA_{a,mm}=0$ for every band $m$, so all
diagonal perturbation matrix elements vanish. For a uniform electric field,
$\alpha^a=-eE^a$, this is the parallel-field condition
$E^aA_{a,mm}=0$.

\begin{figure*}[!t]
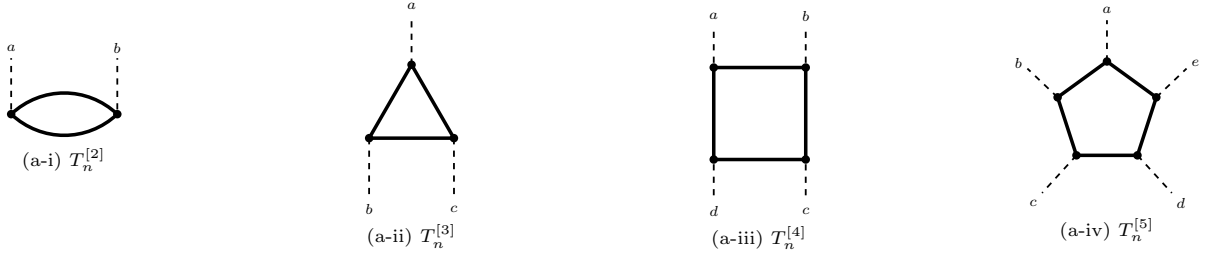

\centering
\begin{minipage}[t]{0.23\textwidth}\centering
\scalebox{0.78}{\EnergyBubbleBW}\\[0.04em]
{\scriptsize (a-i) \(T_n^{[2]}\)}
\end{minipage}\hfill
\begin{minipage}[t]{0.23\textwidth}\centering
\scalebox{0.78}{\EnergyTriangleBW}\\[0.04em]
{\scriptsize (a-ii) \(T_n^{[3]}\)}
\end{minipage}\hfill
\begin{minipage}[t]{0.23\textwidth}\centering
\scalebox{0.78}{\EnergySquareBW}\\[0.04em]
{\scriptsize (a-iii) \(T_n^{[4]}\)}
\end{minipage}\hfill
\begin{minipage}[t]{0.23\textwidth}\centering
\scalebox{0.78}{\EnergyPentagonBW}\\[0.04em]
{\scriptsize (a-iv) \(T_n^{[5]}\)}
\end{minipage}
\caption{BW vertex-order energy diagrams. Thick loop lines denote
intermediate-band sums weighted by dressed resolvents
$1/\Dtilde_{nm}$, and dashed external legs label the perturbation weights
$\alpha^a$.}
\label{fig:vertex-energy-appendix}
\end{figure*}

\begin{figure*}[!t]
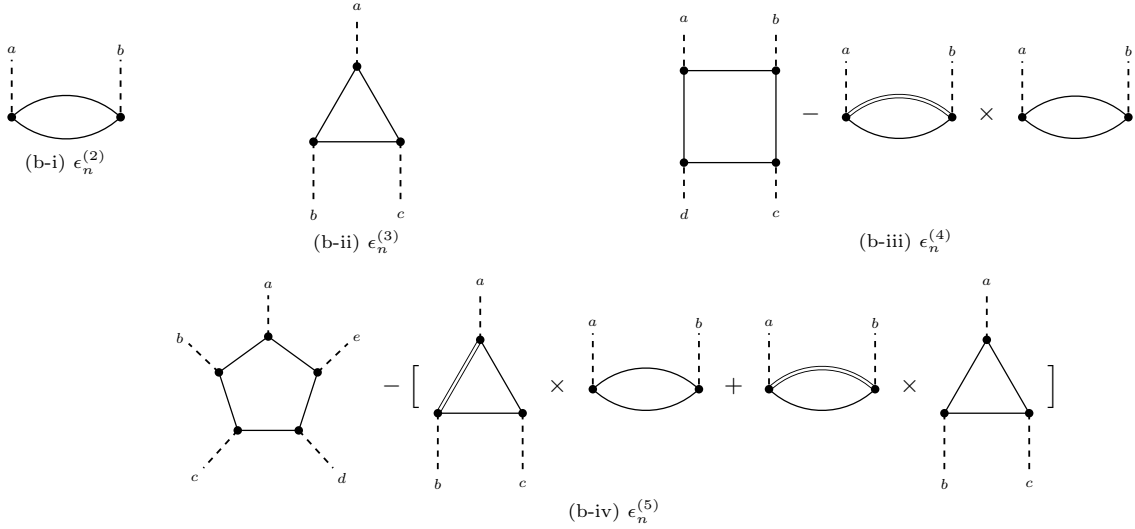

\centering
\begin{minipage}[t]{0.19\textwidth}\centering
\scalebox{0.80}{\EnergyBubbleBare}\\[0.04em]
{\scriptsize (b-i) \(\epsilon_n^{(2)}\)}
\end{minipage}\hfill
\begin{minipage}[t]{0.19\textwidth}\centering
\scalebox{0.80}{\EnergyTriangleBare}\\[0.04em]
{\scriptsize (b-ii) \(\epsilon_n^{(3)}\)}
\end{minipage}\hfill
\begin{minipage}[t]{0.57\textwidth}\centering
\scalebox{0.78}{\EnergySquareBare}
\(\; - \;\)
\scalebox{0.78}{\EnergyBubbleDoubleBare}
\(\;\times\;\)
\scalebox{0.78}{\EnergyBubbleBare}\\[0.04em]
{\scriptsize (b-iii) \(\epsilon_n^{(4)}\)}
\end{minipage}

\vspace{0.65em}

\begin{minipage}[t]{0.98\textwidth}\centering
\scalebox{0.78}{\EnergyPentagonBare}
\(\; - \;\Bigl[\)
\scalebox{0.78}{\EnergyTriangleDoubleBare}
\(\;\times\;\)
\scalebox{0.78}{\EnergyBubbleBare}
\(\; + \;\)
\scalebox{0.78}{\EnergyBubbleDoubleBare}
\(\;\times\;\)
\scalebox{0.78}{\EnergyTriangleBare}
\(\Bigr]\)\\[0.04em]
{\scriptsize (b-iv) \(\epsilon_n^{(5)}\)}
\end{minipage}
\caption{Strict-order energy diagrams through fifth order. Thin lines are
bare resolvents, multiple thin lines denote higher bare pole order, and
products joined by $\times$ are disconnected factors. External-index symmetrization and all allowed
higher-pole placements are understood.}
\label{fig:strict-energy-appendix}
\end{figure*}

The perturbed state in Brillouin--Wigner (BW) form is
\begin{equation}
|\psi_n\rangle
=
\sum_{r\ge0}|\psi_n^{[r]}\rangle,
\qquad
|\psi_n^{[r]}\rangle
=
(R_nV)^r|u_n\rangle,
\label{eq:BWstate}
\end{equation}
with resolvent
\begin{equation}
R_n
=
\sum_{m\ne n}
\frac{|u_m\rangle\langle u_m|}{\Dtilde_{nm}},
\qquad
\Dtilde_{nm}=\epsilon_n-\epsilon_m^{(0)}.
\label{eq:BWresolvent}
\end{equation}
The state is in intermediate normalization, $\langle u_n|\psi_n\rangle=1$, so all
field dependence beyond the reference band resides in the interband
components. Since $V_{nn}=0$, the dressed energy obeys
$\epsilon_n=\epsilon_n^{(0)}+\langle u_n|V|\psi_n\rangle$. The contribution
with $N+1$ perturbation vertices is therefore the dressed closed loop
\begin{equation}
T_n^{[N+1]}
=
\langle u_n|V|\psi_n^{[N]}\rangle
=
\tilde Q^{(N),n}_{(0)\,a_0\cdots a_N}
\alpha^{a_0}\cdots\alpha^{a_N},
\label{eq:appTdef}
\end{equation}
where
\begin{equation}
\begin{split}
\tilde Q^{(N),n}_{(0)\,a_0\cdots a_N}
&=
\cS_{a_0\cdots a_N}
\sum_{\substack{m_1,\ldots,m_N\ne n\\
                  m_j\ne m_{j+1}\;(j=1,\ldots,N-1)}}
\frac{
A_{a_0,nm_1}
\left(\prod_{j=1}^{N-1}A_{a_j,m_jm_{j+1}}\right)
A_{a_N,m_Nn}
}{
\prod_{j=1}^{N}\Dtilde_{nm_j}
}.
\end{split}
\label{eq:appTildeQloop}
\end{equation}
Each summand is the single closed path
$n\to m_1\to m_2\to\cdots\to m_N\to n$.  Every intermediate label lies
outside the reference band, and adjacent labels are unequal because
$V_{mm}=0$. For example, the three-vertex term is
\begin{equation}
\tilde Q^{(2),n}_{(0)\,abc}
=
\cS_{abc}
\sum_{\substack{m,\ell\ne n\\m\ne\ell}}
\frac{A_{a,nm}A_{b,m\ell}A_{c,\ell n}}
{\Dtilde_{nm}\Dtilde_{n\ell}},
\label{eq:appTildeQtriangle}
\end{equation}
which is the triangle path $n\to m\to\ell\to n$.  The bubble, triangle,
square, and pentagon realizations of Eq.~\eqref{eq:appTildeQloop} are shown
in Fig.~\ref{fig:vertex-energy-appendix}.
Vertex order keeps the exact energy $\epsilon_n$ in every BW resolvent.
The corresponding bare-gap tensor, in terms of which the strict-order
expansion is ultimately expressed, is
\begin{equation}
\begin{split}
\bar Q^{(N),n}_{(0)\,a_0\cdots a_N}
&=
\cS_{a_0\cdots a_N}
\sum_{\substack{m_1,\ldots,m_N\ne n\\
                  m_j\ne m_{j+1}\;(j=1,\ldots,N-1)}}
\frac{
A_{a_0,nm_1}
\left(\prod_{j=1}^{N-1}A_{a_j,m_jm_{j+1}}\right)
A_{a_N,m_Nn}
}{
\prod_{j=1}^{N}\Delta_{nm_j}
}.
\end{split}
\label{eq:appBarQloop}
\end{equation}
Thus $\bar Q^{(N)}_{(0)}$ is obtained from
$\tilde Q^{(N)}_{(0)}$ by replacing every dressed gap with its bare
counterpart. The same replacement defines the barred tensor in each
covariant derivative sector, with the pole labels retaining the corresponding
denominator powers.
Strict order in $\alpha$ is obtained by writing
$\epsilon_n=\epsilon_n^{(0)}+\delta\epsilon_n$ and reexpanding each dressed
resolvent in bare gaps,
\begin{equation}
\begin{split}
\tilde G_m^{(n)}
\equiv\frac{1}{\Dtilde_{nm}}
&=
G_m^{(n)}
-
\bigl(G_m^{(n)}\bigr)^2\delta\epsilon_n
+
\bigl(G_m^{(n)}\bigr)^3(\delta\epsilon_n)^2
-\cdots,
\\
G_m^{(n)}&=\frac{1}{\Delta_{nm}},
\qquad
\delta\epsilon_n=\sum_{K\ge1}\epsilon_n^{(K+1)}.
\end{split}
\label{eq:gapExp}
\end{equation}
Inserting Eq.~\eqref{eq:gapExp} into the dressed loop
Eq.~\eqref{eq:appTdef} and matching equal powers of $\alpha$ recursively
expresses the dressed energy entirely in terms of bare gaps as summarized in Fig.~\ref{fig:line-rule}; applying
it recursively to the closed loops of Fig.~\ref{fig:vertex-energy-appendix}
produces Fig.~\ref{fig:strict-energy-appendix}.
\tikzset{
  strictenergyvertex/.style={
    vertex,
    inner sep=1.16pt
  },
  pics/strictenergybubble/.style={
    code={
      \draw[bare]
        (-0.72,0)
        .. controls (-0.38,0.38) and (0.38,0.38) ..
        (0.72,0);
      \draw[bare]
        (-0.72,0)
        .. controls (-0.38,-0.38) and (0.38,-0.38) ..
        (0.72,0);
      \draw[field] (-0.72,0) -- (-0.72,0.68);
      \draw[field] ( 0.72,0) -- ( 0.72,0.68);

      \node[strictenergyvertex] at (-0.72,0) {};
      \node[strictenergyvertex] at ( 0.72,0) {};
    }
  }
}

\begin{figure*}[!t]
\centering
\begin{tikzpicture}[x=1cm,y=1cm]
\draw[bw] (-6.40,1.25) -- (-4.20,1.25);
\node at (-3.70,1.25) {$=$};

\draw[bare] (-3.20,1.25) -- (-1.00,1.25);
\node at (-0.50,1.25) {$-$};

\draw[poleii] (0.00,1.25) -- (2.20,1.25);
\node at (2.70,1.25) {$\times$};

\pic at (4.05,1.25) {strictenergybubble};

\node at (-3.70,-1.25) {$+$};

\draw[poleiii] (-3.20,-1.25) -- (-1.00,-1.25);
\node at (-0.50,-1.25) {$\times$};

\pic at (0.85,-1.25) {strictenergybubble};

\node at (2.05,-1.25) {$\times$};

\pic at (3.40,-1.25) {strictenergybubble};

\node at (4.75,-1.25) {$-\cdots$};
\end{tikzpicture}

\caption{Line-level strict-order expansion. Each closed bubble explicitly represents the leading
strict-order energy correction $\epsilon_n^{(2)}$ substituted for one
factor of $\delta\epsilon_n$. The doubled and tripled thin lines denote
$(G_m^{(n)})^2$ and $(G_m^{(n)})^3$, respectively.}
\label{fig:line-rule}
\end{figure*}
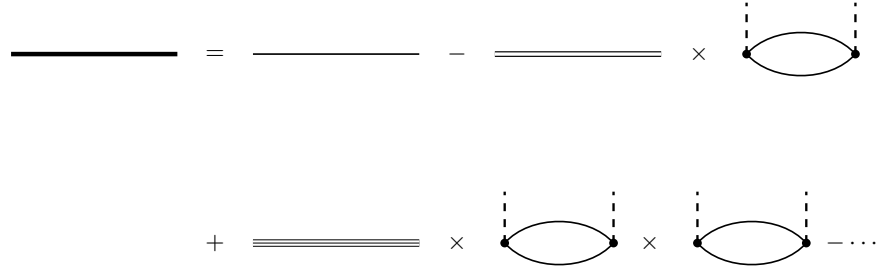
Here the term
$-(G_m^{(n)})^2\delta\epsilon_n$ supplies one extra factor of the same bare resolvent, so that the denominator on that
segment is squared.  Since $\delta\epsilon_n$ is itself a sum of lower-order
strict energy corrections, substituting it recursively factorizes the term
into an energy loop with one higher-pole segment and a separate lower-order
energy loop. The leading strict-order structure is
\begin{align}
\epsilon_n^{(N+1)}
&=
\bar Q^{(N),n}_{(0)}\alpha^{N+1}
\nonumber\\
&\quad-
\sum_{K_1K_2}^{\prime}
\cS\!\left[
\bar Q^{(K_1),n}_{(0)}
\times
\bar Q^{(K_2),n}_{(0)}\big|_{2}
\right]
\alpha^{N+1}
+\cdots .
\label{eq:appStrictE}
\end{align}
In this sum, $K_1$ labels the lower-order energy correction inserted
through $\delta\epsilon_n$, whereas $K_2$ labels the loop containing the
squared resolvent.  The two factors carry orders $\alpha^{K_1+1}$ and
$\alpha^{K_2+1}$, respectively; matching their product to
$\epsilon_n^{(N+1)}\sim\alpha^{N+1}$ gives
$K_1+K_2=N-1$.  The bounds $K_1,K_2\ge1$ follow from $V_{nn}=0$, because the
first nonzero energy correction is second order.  Thus
$(K_1,K_2)=(1,1)$ is the sole subtraction in $\epsilon_n^{(4)}$, while
$(1,2)$ and $(2,1)$ give the two subtractions in $\epsilon_n^{(5)}$.
The pole label $|_2$ marks the segment whose bare denominator is squared,
and $\cS$ includes every allowed placement of that higher pole and every
external-index symmetrization. Repeated terms in
Eq.~\eqref{eq:gapExp} generate the omitted products of three or more
lower-order loops. This is the leading strict-order energy expansion.

\subsection{Berry connection corrections}

\begin{figure*}[!t]
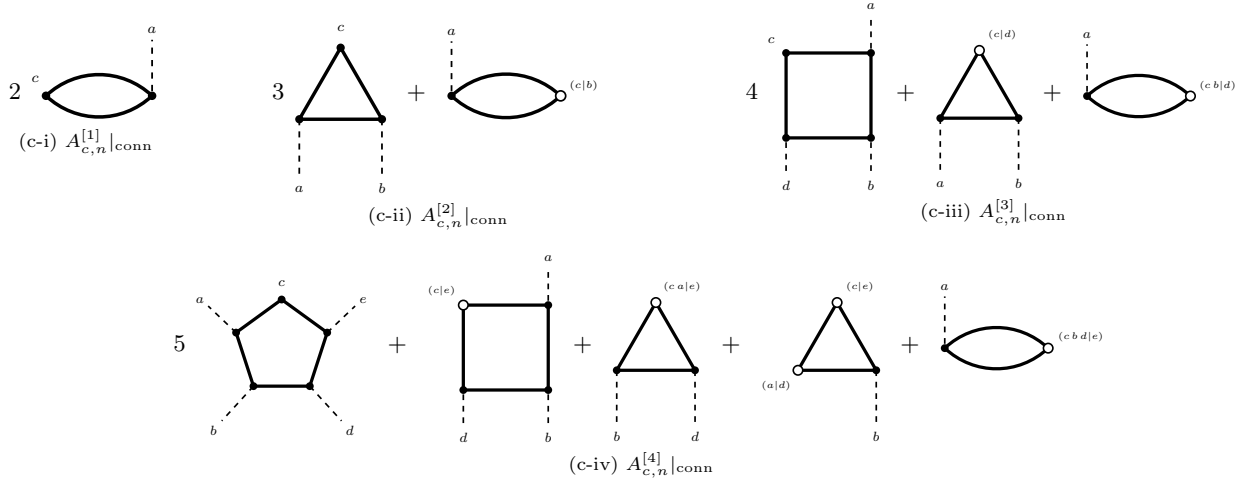

\centering
\begin{minipage}[t]{0.18\textwidth}\centering
\(2\,\)\scalebox{0.78}{\ConnBubbleBW}\\[0.04em]
{\scriptsize (c-i) \(A^{[1]}_{c,n}|_{\conn}\)}
\end{minipage}\hfill
\begin{minipage}[t]{0.30\textwidth}\centering
\(3\,\)\scalebox{0.76}{\ConnTriangleBW}
\(\; + \;\)
\scalebox{0.76}{\ConnCovBubbleBW}\\[0.04em]
{\scriptsize (c-ii) \(A^{[2]}_{c,n}|_{\conn}\)}
\end{minipage}\hfill
\begin{minipage}[t]{0.48\textwidth}\centering
\(4\,\)\scalebox{0.72}{\ConnSquareBW}
\(\; + \;\)
\scalebox{0.72}{\ConnCovTriangleBW}
\(\; + \;\)
\scalebox{0.72}{\ConnCovBubbleTwoBW}\\[0.04em]
{\scriptsize (c-iii) \(A^{[3]}_{c,n}|_{\conn}\)}
\end{minipage}

\vspace{0.70em}

\begin{minipage}[t]{0.98\textwidth}\centering
\(5\,\)\scalebox{0.72}{\ConnPentagonBW}
\(\; + \;\)
\scalebox{0.72}{\ConnCovSquareBW}
\(\; + \;\)
\scalebox{0.72}{\ConnCovTriTwoOneBW}
\(\; + \;\)
\scalebox{0.72}{\ConnCovTriTwoSplitBW}
\(\; + \;\)
\scalebox{0.72}{\ConnCovBubbleThreeBW}\\[0.04em]
{\scriptsize (c-iv) \(A^{[4]}_{c,n}|_{\conn}\)}
\end{minipage}
\caption{Vertex-order connected Berry connection diagrams. Thick loops carry
dressed resolvents. The derivative-free loop at order $N$ has coefficient $N+1$ from the allowed positions of the free connection index.}
\label{fig:vertex-connection-appendix}
\end{figure*}

The Berry connection follows from the same perturbed state after normalization,
\begin{equation}
A_{c,n}(\alpha)
=
-\Imm\,
\frac{\langle\psi_n|\del_c\psi_n\rangle}
{\langle\psi_n|\psi_n\rangle}.
\label{eq:appAalpha}
\end{equation}
To evaluate the numerator at vertex order, we expand the bra and ket chains
on either side of $\del_c$ and collect the order-$N$ term
\begin{equation}
\mathcal{N}^{[N]}_{c,n}
=
-\Imm\sum_{s=0}^{N}
\langle\psi_n^{[s]}|\del_c\psi_n^{[N-s]}\rangle .
\label{eq:appConnectionNumerator}
\end{equation}
Using
$\del_c|u_m\rangle=-i\sum_l|u_l\rangle A_{c,lm}$, a derivative acting on
a band ket or bra joins the two open chains into one closed loop.  The
integer $s$ specifies where that loop is cut into bra and ket chains, and
$s=0,\ldots,N$ gives $N+1$ possible cuts.  After cyclic relabeling and
symmetrization, these terms are identical to the derivative-free loop with
the free index $c$ inserted at any one of its $N+1$ vertices, hence
the $N+1$ coefficient below. Furthermore, when the derivative acts inside an open chain, the ordinary derivative of
an interband connection and the diagonal Berry connections generated by
differentiating its adjacent basis states combine into the covariant block
$D_cA$. Applying the same completeness rearrangement and recursion as in
Sec.~\ref{app:recursion} organizes all such internal terms into the
covariant-derivative sectors $\tilde Q^{(N)}_{(r);c}$ of total derivative
order $1\le r\le N-1$. Finally, since the energy gaps are real, terms in
which $\del_c$ differentiates a resolvent combine into a real contribution
after bra--ket pairing and therefore do not contribute to $-\Imm$.
Consequently the numerator becomes
\begin{equation}
\begin{split}
\mathcal{N}^{[N]}_{c,n}
&=
\tilde Q^{(N),n}_{\conn,c}\alpha^N,
\\
\tilde Q^{(N),n}_{\conn,c}
&=
(N+1)\tilde Q^{(N),n}_{(0);c}
+
\sum_{r=1}^{N-1}\tilde Q^{(N),n}_{(r);c}.
\end{split}
\label{eq:appVertexAconn}
\end{equation}
Here denominators are
still the dressed gaps $\Dtilde_{nm}$.  No bare-gap or strict-order
reexpansion has yet been made.  Moreover, every nontrivial term from the
inverse normalization is a norm loop multiplying a lower-order connection,
so it is disconnected and cannot change the single-loop result above.
For the first two orders this gives
\begin{align}
A^{[1]}_{c,n}\big|_{\conn}
&=2\tilde Q^{(1),n}_{(0);c\,a}\alpha^a,
\nonumber\\
A^{[2]}_{c,n}\big|_{\conn}
&=\left[
3\tilde Q^{(2),n}_{(0);c\,ab}
+\tilde Q^{(2),n}_{(1);c\,ab}\big|_2
\right]\alpha^a\alpha^b.
\label{eq:appConnectionExamples}
\end{align}
These are respectively the bubble and the triangle plus covariant-bubble
structures in Fig.~\ref{fig:vertex-connection-appendix}; the same counting
and covariant differentiation generate the higher panels of that figure. The connected numerator is put in strict order by the same dressed-gap
expansion used for the energy in Eq.~\eqref{eq:gapExp}. \\

\begin{figure*}[!t]
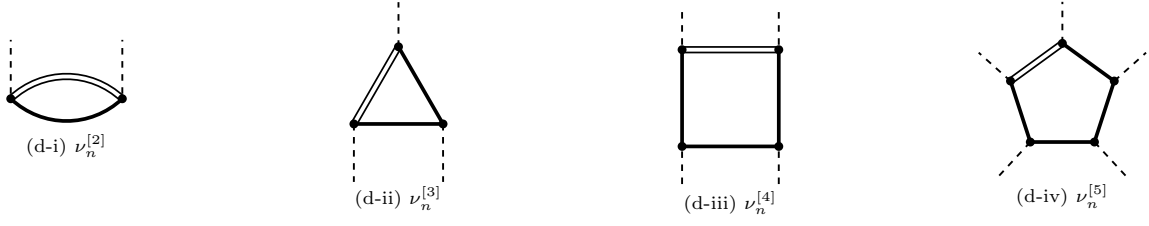

\centering
\begin{minipage}[t]{0.235\textwidth}\centering
\scalebox{0.82}{\NormBubbleBW}\\[0.04em]
{\scriptsize (d-i) \(\nu_n^{[2]}\)}
\end{minipage}%
\hspace{0.01\textwidth}%
\begin{minipage}[t]{0.235\textwidth}\centering
\scalebox{0.82}{\NormTriangleBW}\\[0.04em]
{\scriptsize (d-ii) \(\nu_n^{[3]}\)}
\end{minipage}%
\hspace{0.01\textwidth}%
\begin{minipage}[t]{0.235\textwidth}\centering
\scalebox{0.82}{\NormSquareBW}\\[0.04em]
{\scriptsize (d-iii) \(\nu_n^{[4]}\)}
\end{minipage}%
\hspace{0.01\textwidth}%
\begin{minipage}[t]{0.235\textwidth}\centering
\scalebox{0.82}{\NormPentagonBW}\\[0.04em]
{\scriptsize (d-iv) \(\nu_n^{[5]}\)}
\end{minipage}
\caption{Vertex-order norm diagrams. A diagram assumes a sum over every placement of the squared gap; equivalently every allowed bra--ket cut.}
\label{fig:vertex-norm-appendix}
\end{figure*}

\begin{figure*}[!t]
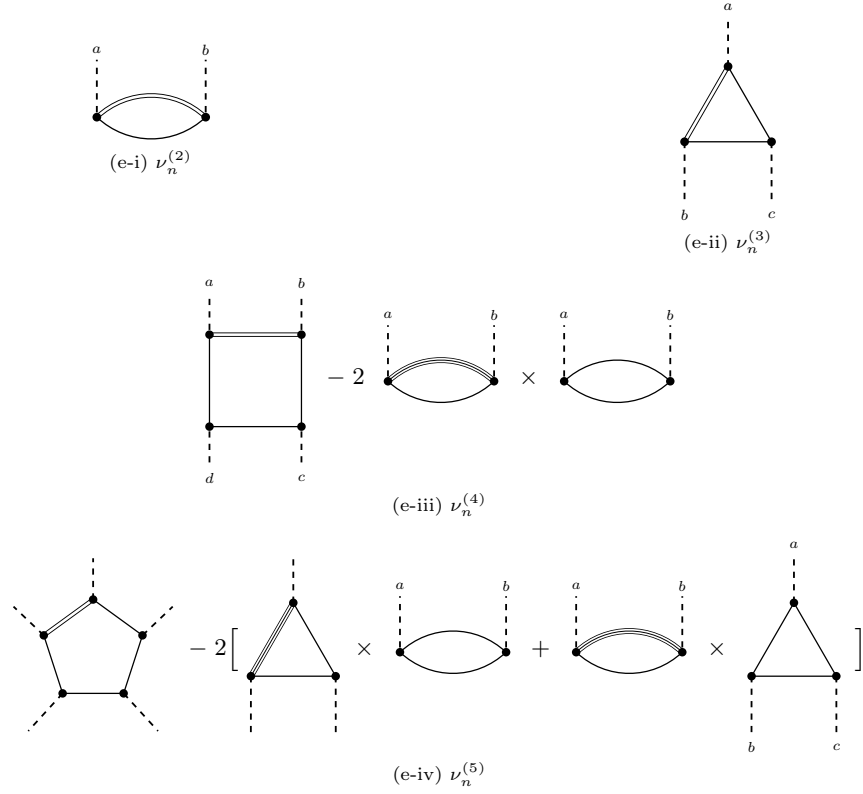

\centering
\begin{minipage}[t]{0.30\textwidth}\centering
\scalebox{0.80}{\EnergyBubbleDoubleBare}\\[0.04em]
{\scriptsize (e-i) \(\nu_n^{(2)}\)}
\end{minipage}\hspace{0.12\textwidth}
\begin{minipage}[t]{0.30\textwidth}\centering
\scalebox{0.80}{\EnergyTriangleDoubleBare}\\[0.04em]
{\scriptsize (e-ii) \(\nu_n^{(3)}\)}
\end{minipage}

\vspace{0.65em}

\begin{minipage}[t]{0.82\textwidth}\centering
\scalebox{0.78}{\EnergySquareDoubleBare}
\(\; - \;2\,\)
\scalebox{0.78}{\EnergyBubbleTripleBare}
\(\;\times\;\)
\scalebox{0.78}{\EnergyBubbleBare}\\[0.04em]
{\scriptsize (e-iii) \(\nu_n^{(4)}\)}
\end{minipage}

\vspace{0.65em}

\begin{minipage}[t]{0.98\textwidth}\centering
\scalebox{0.78}{\EnergyPentagonDoubleBare}
\(\; - \;2\Bigl[\)
\scalebox{0.78}{\EnergyTriangleTripleBare}
\(\;\times\;\)
\scalebox{0.78}{\EnergyBubbleBare}
\(\; + \;\)
\scalebox{0.78}{\EnergyBubbleTripleBare}
\(\;\times\;\)
\scalebox{0.78}{\EnergyTriangleBare}
\(\Bigr]\)\\[0.04em]
{\scriptsize (e-iv) \(\nu_n^{(5)}\)}
\end{minipage}
\caption{Strict-order norm diagrams through fifth order. The factor $-2$ follows from expanding $(\Dtilde)^{-2}$. Each
higher-pole diagram includes the sum over its allowed pole placements.}
\label{fig:strict-norm-appendix}
\end{figure*}

Besides the dressed gap expansion, the normalization denominator supplies the second ingredient needed for
strict order. We write
\begin{equation}
\langle\psi_n|\psi_n\rangle
=
1+\sum_{s\ge2}\nu_n^{[s]},
\qquad
\nu_n^{[s]}
=
\sum_{\substack{r+r'=s\\ r,r'\ge1}}
\langle\psi_n^{[r]}|\psi_n^{[r']}\rangle .
\label{eq:nuDef}
\end{equation}
Using $|\psi_n^{[r]}\rangle=(R_nV)^r|u_n\rangle$, each term in
Eq.~\eqref{eq:nuDef} can be written as
\begin{equation}
\langle\psi_n^{[r]}|\psi_n^{[s-r]}\rangle
=
\langle u_n|
(VR_n)^r(R_nV)^{s-r}
|u_n\rangle .
\label{eq:nuChains}
\end{equation}
The last resolvent of the bra chain and the first resolvent of the ket
chain act on the same intermediate state. They therefore meet at the
bra--ket cut as $
R_nR_n=R_n^2$. Consequently, every contribution to $\nu_n^{[s]}$ is a single closed
loop with $s$ perturbation vertices, one squared dressed resolvent at
the bra--ket cut, and ordinary dressed resolvents on the remaining
virtual segments. For instance, the first two orders are
\begin{align}
\nu_n^{[2]}
&=
\langle u_n|VR_n^2V|u_n\rangle ,
\label{eq:nuTwoExplicit}\\
\nu_n^{[3]}
&=
\langle u_n|VR_n^2VR_nV|u_n\rangle
+
\langle u_n|VR_nVR_n^2V|u_n\rangle .
\label{eq:nuThreeExplicit}
\end{align}
Thus the partition $r+(s-r)=s$ specifies the position of the
bra--ket cut and hence the position of the squared resolvent. The sum
over $r=1,\ldots,s-1$, together with the symmetrization over external
perturbation indices, is represented diagrammatically by one loop with all allowed squared-pole placements understood. The
bubble, triangle, square, and pentagon in
Fig.~\ref{fig:vertex-norm-appendix} are precisely the resulting
$\nu_n^{[2]}$, $\nu_n^{[3]}$, $\nu_n^{[4]}$, and $\nu_n^{[5]}$ loops.
Expanding the inverse norm in Eq.~\eqref{eq:appAalpha} multiplies
lower-order connection corrections by norm loops. The boundary
term $-\nu_n^{[N]}A^{(0)}_{c,n}$ is gauge dependent and cancels identically
against the diagonal part of the numerator at the same order,
$A^{(0)}_{c,n}\,\Ree\langle\psi_n^{[r]}|\psi_n^{[r']}\rangle$ with
$r+r'=N$. Consequently, norm loops multiply only genuinely perturbative
connection corrections,
\begin{equation}
A_{c,n}^{[N]}\big|_{\rm dis}
=
-\!\sum_{s=2}^{N-1}\nu_n^{[s]}A_{c,n}^{[N-s]}
+
\!\!\!\sum_{\substack{s,s'\ge2\\ s+s'\le N-1}}\!\!\!
\nu_n^{[s]}\nu_n^{[s']}A_{c,n}^{[N-s-s']}
-\cdots .
\label{eq:Aret}
\end{equation}
For the norm diagrams, the dressed denominator is squared,
so its bare-gap expansion is
\begin{equation}
(\Dtilde_{nm})^{-2}
=
\bigl(G_m^{(n)}\bigr)^2
-
2\delta\epsilon_n
\bigl(G_m^{(n)}\bigr)^3
+\cdots .
\label{eq:normPoleReexp}
\end{equation}
Applying Eq.~\eqref{eq:normPoleReexp} to the vertex-order norm loops of
Fig.~\ref{fig:vertex-norm-appendix} gives the bare-pole norm diagrams in
Fig.~\ref{fig:strict-norm-appendix}. \\
\\

\begin{figure*}[!t]
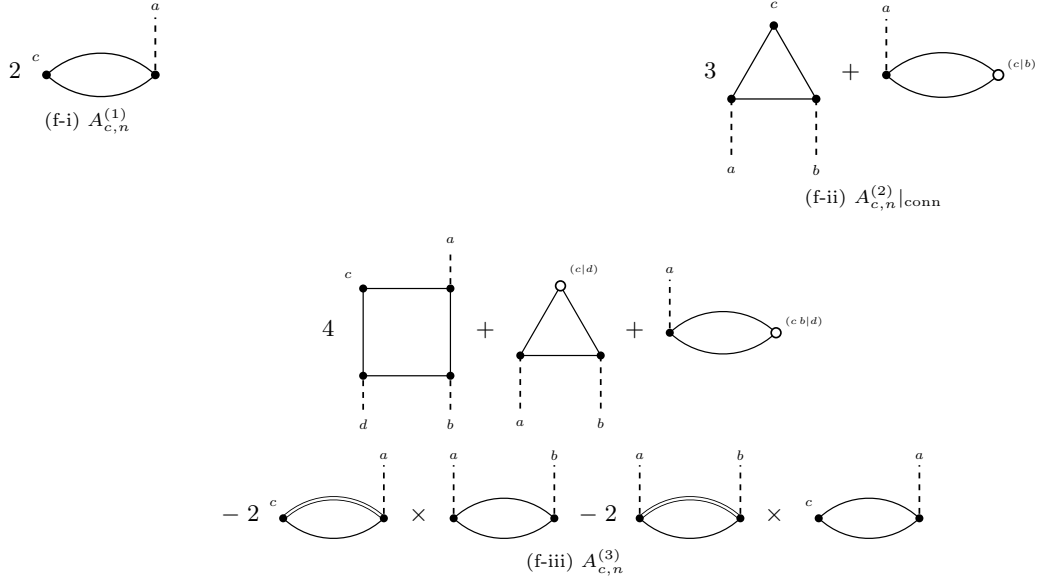

\centering
\begin{minipage}[t]{0.28\textwidth}\centering
\(2\,\)\scalebox{0.80}{\ConnBubbleBare}\\[0.04em]
{\scriptsize (f-i) \(A^{(1)}_{c,n}\)}
\end{minipage}\hfill
\begin{minipage}[t]{0.56\textwidth}\centering
\(3\,\)\scalebox{0.78}{\ConnTriangleBare}
\(\; + \;\)
\scalebox{0.78}{\ConnCovBubbleBare}\\[0.04em]
{\scriptsize (f-ii) \(A^{(2)}_{c,n}|_{\conn}\)}
\end{minipage}

\vspace{0.70em}

\begin{minipage}[t]{0.98\textwidth}\centering
\(4\,\)\scalebox{0.74}{\ConnSquareBare}
\(\; + \;\)
\scalebox{0.74}{\ConnCovTriangleBare}
\(\; + \;\)
\scalebox{0.74}{\ConnCovBubbleTwoBare}

\vspace{0.28em}

\( -\;2\,\)\scalebox{0.74}{\ConnBubbleDoubleBare}
\(\;\times\;\)
\scalebox{0.74}{\EnergyBubbleBare}
\(\; - \;2\,\)
\scalebox{0.74}{\EnergyBubbleDoubleBare}
\(\;\times\;\)
\scalebox{0.74}{\ConnBubbleBare}\\[0.04em]
{\scriptsize (f-iii) \(A^{(3)}_{c,n}\)}
\end{minipage}
\caption{Strict-order Berry connection diagrams through third order. The first
line of panel (f-iii) is the bare connected correction. The two subtractions are
the dressed-gap correction and the normalization
correction, respectively. Their coefficient $2$ is inherited from the two placements of
the free connection index in $A^{(1)}_{c,n}$.}
\label{fig:strict-connection-appendix-low}
\end{figure*}

\begin{figure*}[!t]
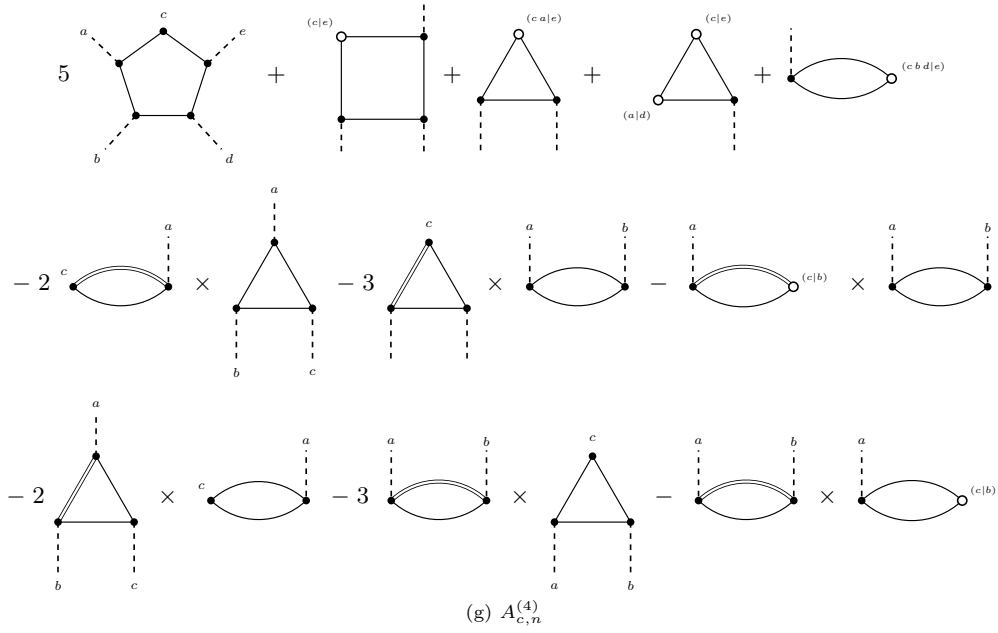

\centering
\begin{minipage}[t]{0.98\textwidth}\centering
\(5\,\)\scalebox{0.70}{\ConnPentagonBare}
\(\; + \;\)
\scalebox{0.70}{\ConnCovSquareBare}
\(\; + \;\)
\scalebox{0.70}{\ConnCovTriTwoOneBare}
\(\; + \;\)
\scalebox{0.70}{\ConnCovTriTwoSplitBare}
\(\; + \;\)
\scalebox{0.70}{\ConnCovBubbleThreeBare}

\vspace{0.42em}

\( -\;2\,\)\scalebox{0.70}{\ConnBubbleDoubleBare}
\(\;\times\;\)
\scalebox{0.70}{\EnergyTriangleBare}
\(\; - \;3\,\)
\scalebox{0.70}{\ConnTriangleDoubleBare}
\(\;\times\;\)
\scalebox{0.70}{\EnergyBubbleBare}
\(\; - \;\)
\scalebox{0.70}{\ConnCovBubbleHigherPoleBare}
\(\;\times\;\)
\scalebox{0.70}{\EnergyBubbleBare}

\vspace{0.42em}

\( -\;2\,\)\scalebox{0.70}{\EnergyTriangleDoubleBare}
\(\;\times\;\)
\scalebox{0.70}{\ConnBubbleBare}
\(\; - \;3\,\)
\scalebox{0.70}{\EnergyBubbleDoubleBare}
\(\;\times\;\)
\scalebox{0.70}{\ConnTriangleBare}
\(\; - \;\)
\scalebox{0.70}{\EnergyBubbleDoubleBare}
\(\;\times\;\)
\scalebox{0.70}{\ConnCovBubbleBare}\\[0.04em]
{\scriptsize (g) \(A^{(4)}_{c,n}\)}
\end{minipage}
\caption{Complete strict-order connection diagrams at fourth order. The
first row is the bare connected correction, the second row contains the
dressed-gap subtractions, and the third row contains the normalization
subtractions. The factors $2$ and $3$ are inherited from the derivative-free
parts of $A^{(1)}$ and $A^{(2)}$. Higher-pole symbols include all allowed pole
placements. Together with the
strict-order energy diagrams through fifth order in
Fig.~\ref{fig:strict-energy-appendix}, these diagrams provide the remaining
ingredients needed to construct the fifth-order charge conductivity.}
\label{fig:strict-connection-appendix-fourth}
\end{figure*}

At this point we have everything needed to put together the strict-order Berry connection corrections. Combining gives
\begin{align}
A_{c,n}^{(N)}
&=
\bar Q^{(N),n}_{\conn,c}\alpha^N
\nonumber\\
&\quad-
\sum_{K_1K_2}^{\prime}
\cS\!\left[
\bar Q^{(K_1),n}_{\conn,c}\big|_{2}
\bar Q^{(K_2),n}_{(0)}
+
\bar Q^{(K_1),n}_{\conn,c}
\bar Q^{(K_2),n}_{(0)}\big|_{2}
\right]
\alpha^N
+\cdots .
\label{eq:appStrictA}
\end{align}
Here $K_1$ is the rank of the perturbative connection factor, which
starts at order $\alpha^{K_1}$, and $K_2$ is the rank of the energy or norm
loop, which starts at order $\alpha^{K_2+1}$. Their product therefore has
order $\alpha^{K_1+K_2+1}$, so matching $A_{c,n}^{(N)}$ gives
$K_1+K_2=N-1$. The earlier gauge-dependent term cancellation enforces the lower bounds $K_1,K_2\ge1$. The first product is the
dressed-gap subtraction in a connected connection loop. The second is the
normalization subtraction, in which the connection correction is multiplied
by a norm loop whose leading bare form is an energy loop with one squared
pole. The numerical weights are inherited from
Eq.~\eqref{eq:appVertexAconn}: the derivative-free connection topology of
rank $K_1$ carries $K_1+1$, whereas the displayed covariant branches carry
unit weight. These factors are written explicitly in
Figs.~\ref{fig:strict-connection-appendix-low} and
\ref{fig:strict-connection-appendix-fourth}. Higher terms contain repeated gap insertions, products of
several norm loops, and their mixed products. Equation~\eqref{eq:appStrictA}
is the leading strict-order Berry-connection expansion.

\section{Semiclassical transport}
\label{app:transportDerivation}
For a uniform electric field,
\begin{equation}
\hbar \dot k_a=-eE_a,
\qquad
v_{d,n}(\EE)
=
\frac{1}{\hbar}\del_d\epsilon_n(\EE)
-
\frac{e}{\hbar}E_a\Omega^n_{ad}(\EE).
\label{eq:appSemiclassicalEOM}
\end{equation}
The charge current is
\begin{equation}
j_d
=
-e\sum_n\int_{\BZ}\frac{d^d k}{(2\pi)^d}\,
f_n(\kk;\EE)\,v_{d,n}(\kk;\EE),
\label{eq:appCurrentDef}
\end{equation}
where $f_n(\kk;\EE)$ is the nonequilibrium distribution and
$f_n^{(0)}(\kk)$ is the equilibrium Fermi--Dirac distribution evaluated
with the unperturbed energy $\epsilon_n^{(0)}$. The derivation is valid in
arbitrary dimension $d$; all tensor indices range over the corresponding
momentum directions. The field enters through
both the distribution function and the dressed band data,
\begin{align}
\epsilon_n(\EE)
&=
\epsilon_n^{(0)}+
\epsilon_n^{(2)}+
\epsilon_n^{(3)}+
\cdots,
\label{eq:appEpsLowTransport}\\
\Omega^n_{cd}(\EE)
&=
\Omega^{(0),n}_{cd}+
\Omega^{(1),n}_{cd}+
\Omega^{(2),n}_{cd}+
\cdots .
\label{eq:appOmegaLowTransport}
\end{align}
In the relaxation-time approximation and in the dc limit
$\omega\ll1/\tau$, the order-$\ell$ correction to the
distribution is
\begin{equation}
f_n^{(\ell)}
=
\left(\frac{e\tau}{\hbar}\right)^\ell
E_{a_1}\cdots E_{a_\ell}
\del_{a_1}\cdots\del_{a_\ell}f_n^{(0)} .
\label{eq:RTA}
\end{equation}
At arbitrary order $r$, collecting equal powers of the field gives
\begin{equation}
j_d^{(r)}
=-e\sum_n\int_{\BZ}\frac{d^d k}{(2\pi)^d}
\sum_{\ell=0}^{r}
f_n^{(\ell)}v_{d,n}^{(r-\ell)}.
\label{eq:jrGeneralApp}
\end{equation}
In particular, the second-order current is
\begin{equation}
\begin{split}
j_d^{(2)}=-e\sum_n\int_{\BZ}\frac{d^d k}{(2\pi)^d}
\Big[&
 f_n^{(2)}v^{(0)}_{d,n}
+f_n^{(1)}v^{(1)}_{d,n}  \\
&+f_n^{(0)}v^{(2)}_{d,n}
\Big],
\end{split}
\label{eq:j2decompApp}
\end{equation}
while the third-order current is
\begin{equation}
\begin{split}
j_d^{(3)}=-e\sum_n\int_{\BZ}\frac{d^d k}{(2\pi)^d}
\Big[&
 f_n^{(3)}v^{(0)}_{d,n}
+f_n^{(2)}v^{(1)}_{d,n}  \\
&+f_n^{(1)}v^{(2)}_{d,n}
+f_n^{(0)}v^{(3)}_{d,n}
\Big] .
\end{split}
\label{eq:j3decompApp}
\end{equation}
The velocity pieces entering these expressions are
\begin{align}
v^{(0)}_{d,n}
&=\frac{1}{\hbar}\del_d\epsilon_n^{(0)},
\label{eq:v0App}\\
v^{(1)}_{d,n}
&=-\frac{e}{\hbar}E_a\Omega^{(0),n}_{ad},
\label{eq:v1App}\\
v^{(2)}_{d,n}
&=
\frac{1}{\hbar}\del_d\epsilon_n^{(2)}
-
\frac{e}{\hbar}E_a\Omega^{(1),n}_{ad},
\label{eq:v2App}\\
v^{(3)}_{d,n}
&=
\frac{1}{\hbar}\del_d\epsilon_n^{(3)}
-
\frac{e}{\hbar}E_a\Omega^{(2),n}_{ad}.
\label{eq:v3App}
\end{align}
In the parallel-transport gauge specified above, the first-order energy
correction vanishes, $\epsilon_n^{(1)}=V_{nn}=0$, so it does not appear in
$v^{(1)}$. The geometric perturbation theory gives
\begin{align}
\epsilon_n^{(2)}
&=e^2E_aE_b\bar Q^{(1),n}_{(0)ab},
&
\epsilon_n^{(3)}
&=-e^3E_aE_bE_c\bar Q^{(2),n}_{(0)abc},
\label{eq:appEpsCorrections}\\
\Omega^{(1),n}_{cd}
&=-2eE_a
\left(
\del_c\bar Q^{(1),n}_{(0)da}
-
\del_d\bar Q^{(1),n}_{(0)ca}
\right),
&
\Omega^{(2),n}_{cd}
&=e^2E_aE_b\cW^{(2),n}_{cd;ab}.
\label{eq:appOmegaCorrections}
\end{align}
Here $\bar Q^{(1),n}_{(0)ab}$ is the real, field-symmetric,
gap-weighted projection of the rank-two tensor, and
\begin{equation}
\cW^{(2),n}_{cd;ab}
\equiv
\del_c\bar Q^{(2),n}_{\conn,d;ab}
-
\del_d\bar Q^{(2),n}_{\conn,c;ab}.
\label{eq:appWTwo}
\end{equation}
The last identification uses the fact that the second-order connection is
purely connected, the disconnected sector of Eq.~\eqref{eq:Aret} being empty
at this order. We use the equilibrium Fermi-sea bracket
\begin{equation}
\langle X\rangle_f
\equiv
\sum_n\int_{\BZ}\frac{d^d k}{(2\pi)^d}\,
f_n^{(0)}(\kk)X_n(\kk).
\label{eq:appFermiSeaBracket}
\end{equation}
At $r=2$, substitution of
Eqs.~\eqref{eq:RTA}--\eqref{eq:appOmegaCorrections} into
Eq.~\eqref{eq:j2decompApp} and integration by parts give the $\tau^2$,
$\tau$, and $\tau^0$ channels listed below.
At $r=3$, the same steps applied to Eq.~\eqref{eq:j3decompApp} produce the
five third-order channels listed below. Integration by parts turns the distribution derivatives into the average in
Eq.~\eqref{eq:appFermiSeaBracket}.
Suppressing the superscript $(0)$ on the unperturbed band quantities, the
$f^{(2)}v^{(0)}$ term gives
\begin{equation}
\sigma_{D}^{ab;c}
=-
\frac{e^3\tau^2}{\hbar^3}
\left\langle
\del_a\del_b\del_c\epsilon_n
\right\rangle_f,
\label{eq:appSigma2Drude}
\end{equation}
and $f^{(1)}v^{(1)}$ gives
\begin{equation}
\sigma_{\rm BCD}^{ab;c}
=
-
\frac{e^3\tau}{2\hbar^2}
\left\langle
\del_a\Omega^n_{bc}+\del_b\Omega^n_{ac}
\right\rangle_f.
\label{eq:appSigma2BCD}
\end{equation}
Finally, the two pieces of $f^{(0)}v^{(2)}$, from $\epsilon^{(2)}$ and
$\Omega^{(1)}$, combine into
\begin{equation}
\begin{split}
\sigma_{\rm QMD}^{ab;c}
&=-
\frac{e^3}{\hbar}
\Big\langle
\del_a\bar Q^{(1),n}_{(0)bc}
+
\del_b\bar Q^{(1),n}_{(0)ac}
\\[-2pt]
&\hspace{8.0em}
-
\del_c\bar Q^{(1),n}_{(0)ab}
\Big\rangle_f.
\end{split}
\label{eq:appSigma2QMD}
\end{equation}
Equations~\eqref{eq:appSigma2Drude}--\eqref{eq:appSigma2QMD} are the three
second-order channels.\\

Similarly, the term $f^{(3)}v^{(0)}$ gives the intraband Drude channel,
\begin{equation}
\sigma_D^{abc;d}
=
\frac{e^4\tau^3}{\hbar^4}
\left\langle
\del_a\del_b\del_c\del_d\epsilon_n
\right\rangle_f .
\label{eq:appSigmaD}
\end{equation}
The anomalous velocity in $f^{(2)}v^{(1)}$ gives the Berry curvature
quadrupole,
\begin{equation}
\begin{split}
\sigma_{\rm BCQ}^{abc;d}
&=
\frac{e^4\tau^2}{3\hbar^3}
\Big\langle
\del_a\del_b\Omega^n_{cd}
+
\del_b\del_c\Omega^n_{ad}
+
\del_a\del_c\Omega^n_{bd}
\Big\rangle_f .
\end{split}
\label{eq:appSigmaBCQ}
\end{equation}
The $f^{(1)}v^{(2)}$ contribution contains both the band-energy shift
$\epsilon^{(2)}$ and the first curvature correction $\Omega^{(1)}$.
Using Equations~\eqref{eq:appEpsCorrections}--\eqref{eq:appOmegaCorrections} gives
\begin{equation}
\begin{split}
\sigma_{\rm QMQ}^{abc;d}
&=
\frac{e^4\tau}{3\hbar^2}
\Big\langle
\sum_{\mathrm{cyc}(a,b,c)}
\Big[
2\del_a\del_b\bar Q^{(1),n}_{(0)cd}
\\[-2pt]
&\hspace{8.0em}
-\del_a\del_d\bar Q^{(1),n}_{(0)bc}
\Big]
\Big\rangle_f .
\end{split}
\label{eq:appSigmaQMQ}
\end{equation}
The two intrinsic channels come from $f^{(0)}v^{(3)}$. The anomalous
velocity with $\Omega^{(2)}$ yields
\begin{equation}
\sigma_{\cW}^{abc;d}
=
\frac{e^4}{3\hbar}
\left\langle
\cW^{(2),n}_{ad;bc}
+
\cW^{(2),n}_{bd;ac}
+
\cW^{(2),n}_{cd;ab}
\right\rangle_f,
\label{eq:appSigmaW}
\end{equation}
while the group velocity from $\epsilon^{(3)}$ yields
\begin{equation}
\sigma_{\cL}^{abc;d}
=
\frac{e^4}{\hbar}
\left\langle
\del_d\bar Q^{(2),n}_{(0)abc}
\right\rangle_f .
\label{eq:appSigmaL}
\end{equation}
For a smooth periodic tensor defined over the Brillouin zone, integration by parts gives
\begin{equation*}
\left\langle
\del_d\bar Q^{(2),n}_{(0)abc}
\right\rangle_f
=-
\sum_n\int_{\BZ}\frac{d^d k}{(2\pi)^d}\,
(\del_d f_n^{(0)})\bar Q^{(2),n}_{(0)abc}.
\end{equation*}
Thus $\sigma_{\cL}$ is Fermi-surface controlled and vanishes for a completely
filled band. Equations~\eqref{eq:appSigmaD}--\eqref{eq:appSigmaL} are the five
third-order channels.

\end{document}